\documentclass[preprint2]{proto}
\usepackage{times}

\usepackage{natbib}
\bibpunct{(}{)}{;}{a}{,}{,}
\usepackage{graphicx}
\usepackage{pstricks}
\usepackage{hyperref}
\usepackage{breakurl}

\usepackage{etoolbox}
\makeatletter
\patchcmd{\NAT@test}{\else\NAT@nm}{\else\NAT@nmfmt{\NAT@nm}}{}{}
\let\NAT@up\itshape
\makeatother

\renewcommand{\micron}{\ensuremath{\mu}m}

\begin{document}

\title{\textbf{\LARGE Asteroid Models from Multiple Data Sources}}

\author {\textbf{\large Josef \v{D}urech}}
\affil{\small\em Astronomical Institute, Faculty of Mathematics and Physics, Charles University in Prague}
\author {\textbf{\large Beno\^it Carry}}
\affil{\small\em Institut de M{\'e}canique C{\'e}leste et de Calcul des {\'E}ph{\'e}m{\'e}rides}
\author {\textbf{\large Marco Delbo}}
\affil{\small\em Laboratoire Lagrange, UNS-CNRS-Observatoire de la  C\^ote d'Azur}
\author {\textbf{\large Mikko Kaasalainen}}
\affil{\small\em Tampere University of Technology}
\author {\textbf{\large Matti Viikinkoski}}
\affil{\small\em Tampere University of Technology}

\begin{abstract}
\begin{list}{ } {\rightmargin 1in}
\baselineskip = 11pt
\parindent=1pc
{\small In the past decade, hundreds of asteroid shape models have
  been derived using the lightcurve inversion method. At the same
  time, a new framework of 3-D shape modeling based on the combined
  analysis of widely different data sources such as optical
  lightcurves, disk-resolved images, stellar occultation timings,
  mid-infrared thermal radiometry, optical interferometry, and radar
  delay-Doppler data, has been developed. This multi-data approach
  allows the determination of most of the physical and surface
  properties of asteroids in a single, coherent inversion, with
  spectacular results. We review the main results of asteroid
  lightcurve inversion and also recent advances in multi-data
  modeling. We show that models based on remote sensing data were confirmed
  by spacecraft encounters with asteroids, and we discuss how the
  multiplication of highly detailed 3-D models will help to refine our
  general knowledge of the asteroid population. The physical and
  surface properties of asteroids, i.e., their spin, 3-D shape,
  density, thermal inertia, surface roughness, are among the least
  known of all asteroid properties. Apart for the albedo and diameter,
  we have access to the whole picture for only a few hundreds of
  asteroids. These quantities are nevertheless very important to
  understand as they affect the non-gravitational
  Yarkovsky effect responsible for meteorite delivery to Earth, or
  the bulk composition and internal structure of asteroids. 
 \\~\\~\\~}

\end{list}
\end{abstract}

\section{\textbf{INTRODUCTION}}
 
  The determination of asteroid physical properties is an essential
    part of the complex process of revealing the nature of the
    asteroid population.
  In many cases, this process starts with obtaining
  observational data, continues with 
  creating a model of the asteroid
  (i.e., its size, 3-D shape, and spin state, in the first approximation),
  and ends with interpreting new facts based on the model or a set
  of these. 
  In this sense, modeling is a crucial mid-step between
  observations and theory. Results based on individual well-studied
  asteroids can be generalized to other members of the population. On
  the other hand, a statistically large sample of asteroids with known
  properties can reveal physical effects that play an important role for
  the whole population. 
  
  In this chapter, we will build on the
    content of the {\em Asteroids
    III} chapter by \cite{Kaa.ea:02c} about asteroid models
  reconstructed from {\em disk-integrated} photometry. Although visual
  photometry still remains the most important data source of the modeling, the
  main progress in this field since {\em Asteroids III} has been the
  addition of {\em complementary data sources}. Many of these data sources are {\em disk-resolved}, thus containing much more information than disk-integrated data. This  shift in
  paradigm -- using photometry not alone but simultaneously with
  complementary data -- was mentioned in the last paragraph of the
  {\em Asteroids III} chapter as ``{\em perhaps the most interesting future
  prospect\/}'', and we are now at this stage. In the following, we will review all data types suitable
  for inversion, their sources, uncertainties, and how they
  can be used in modeling.  

  When describing the methods of data inversion and the results
  obtained by these methods, it is also important to emphasize
  caveats, ambiguities, and possible sources of errors. Although the
  description of what can be obtained from different data sources is
  exciting, the knowledge of what cannot, i.e., what are
  limitations of 
  our data sets, is of the same importance. Omitting this may lead to
  over interpretation of results. 

  This chapter is structured as follows. First we review
  the main principles of the multimodal inverse problem in Sect.~\ref{sec:inversion}. Then, in Sect.~\ref{sec:data}, we discuss each data
  type and their contribution to model characteristics and details, and we describe some extensions of the predominant model. In
  Sect.~\ref{sec:results}, we discuss the main results based on
  lightcurve inversion and multimodal asteroid reconstruction. 
  We conclude with prospects for the future in Sect.~\ref{sec:future}. 

\section{\textbf{THEORETICAL ASPECTS OF INVERSION AND DATA FUSION}} \label{sec:inversion}

Asteroid physical model reconstruction from multimodal data is, by its nature, a mathematical inverse problem. It is ill-posed; i.e., the uniqueness and stability properties of the solution are usually not very good unless the data are supported by a number of prior constraints. Furthermore, it is not sufficient just to fit some model to the data numerically and try to probe the solution space with some scheme. Although there are more approaches to the problem of asteroid shape reconstruction, they are usually dealing with only one data type and we mention them in the next section. Here, we describe the problem in a general way in the framework of \emph{generalized projections}: our data are various 1-D or 2-D projection types of a 3-D model, and understanding the fundamental mathematical properties of the inverse projection mapping is essential. This includes a number of theorems on uniqueness, information content, and stability properties \citep{Kaa.Lam:06, Kaa:11, Vii.Kaa:14}.

Let the projection point $\xi_0$ in the image plane (plane-of-sky or range-Doppler) of the point $\mathbf{x}_0$ on the body be mapped by the matrix $A$: $\xi_0=A\mathbf{x}_0$. Define the set $\mathcal{I}(\xi)$ for any $\xi$ as 
\begin{equation}
\mathcal{I}(\xi)=\{\mathbf{x}\vert g(\xi,\mathbf{x};R,t)\,h(\mathbf{x};M,R,t)=1\},
\end{equation}
  where we have explicitly shown the time $t$ and the adjustable
  parameters: $M$ for the shape and $R$ for the rotation. The
  projection point function $g(\xi,\mathbf{x})=1$ if
  $A(R,t)\mathbf{x}=\xi$, and $g=0$ otherwise. The ray-tracing
  function $h=1$ if $\mathbf{x}$ is visible (for occultation, 
  thermal, and radar data), or visible and illuminated (for disk-resolved imaging
  and photometry in the optical); otherwise 
  $h=0$.
  The set $\mathcal{I}(\xi)$ is numerable and finite.
  The number of elements in $\mathcal{I}(\xi)$ is at most one for
  plane-of-sky projections (each point on the projection corresponds
  to at most one point of the asteroid's surface); for range-Doppler,
  it can be more (more points on asteroid's surface can have the same
  distance to the observer and the same relative radial
  velocity). Generalized projections, i.e., all the data
  modes presented in Sect.~\ref{sec:data}, can
  now be presented as scalar values $p(\xi)$ in the image field
  $\Omega$: 
\begin{equation}
p(\xi;t)=\int_\Omega f(\xi,\eta)\, \sum_{\mathbf{x}\in\mathcal{I}(\eta)}
S(\mathbf{x};M,R,L,t)\, \mathrm{d}\eta,
\end{equation}
where $L$ denotes the luminosity parameters (for scattering or thermal
properties), and the luminosity function is denoted by $S$. The
function $f$ is the point-spread, pixellation, or other transfer
function of the image field. For interferometry, it is typically the
Fourier transform kernel. In fact, the reconstruction process works
efficiently by taking the Fourier transform of any image type rather
than using the original pixels \citep{Vii.Kaa:14}. For lightcurves,
$f=1$ (and $\xi$ is irrelevant, $p(\xi)$ is constant). The surface
albedo is usually assumed to be constant, although its variegation can
be included in $S$ by the parameters $L$ if there are high-quality
disk-resolved data. In the case of lightcurves only, we can get an
indication of non-uniform albedo and compensate for this with a
(non-unique) spot model \citep{Kaa.ea:01}. 

The multimodal inverse problem can be expressed as follows. Let us
choose as goodness-of-fit measures some functions $\delta_i$,
$i=1,\ldots,n$, of $n$ data modalities.
Typically, $\delta$ is the
usual $\chi^2$-fit form between $p_{\rm model}$ and $p_{\rm obs}$. Our
task is to construct a joint $\delta_{\rm tot}$ with weighting for
each data mode: 
\begin{equation}
\delta_{\rm tot}(P,D)=\delta_1(P,D_1)+\sum_{i=2}^n \lambda_{i-1}
\delta_i(P,D_i),\quad D=\{D_i\},
\end{equation}
where $D_i$ denotes the data from the source $i$, $\lambda_{i-1}$ is the weight of the source $i$, and $P=\{M,R,L\}$ is
the set of model parameter values. The best-fit result is obtained by
minimizing $\delta_{\rm tot}$ with nonlinear techniques, typically
Levenberg-Marquardt for efficient convergence.
Regularization
functions $r(P)$ can be added to the sum; these constrain, for
  instance, the
smoothness of the surface to suppress large variations at small
scales, the deviation from principal-axis rotation to force the model
to rotate around the shortest inertia axis (assuming uniform density), or
the gravitational slope, etc. \citep{Kaa.Vii:12}. 

\begin{figure}[t]
 \plotone{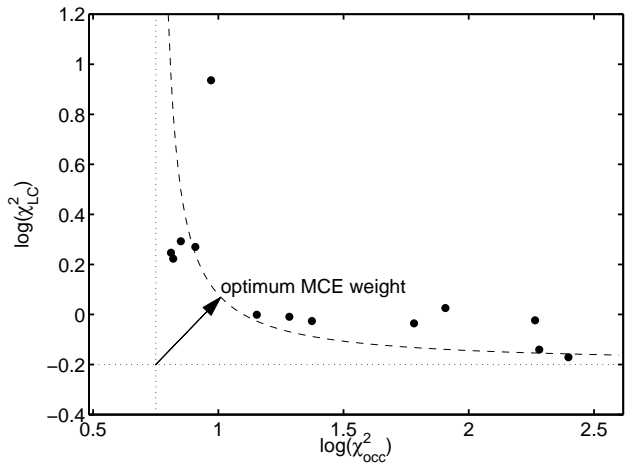}
 \caption{\small The level of fit for lightcurves and occultation data for
   different weighting between the two data types. The optimum weight
   is around the ``corner'' of the L-curve. Each dot corresponds
     to an inversion with a different weight $\lambda$.
 }   
 \label{fig_Hertha_occ_weight}
\end{figure}

The modality (and regularization) weights $\lambda_i$ are determined
using the maximum compatibility estimate (MCE) principle
\citep{Kaa:11,Kaa.Vii:12}.
This yields well-defined unique values that
are, in essence, the best compromise between the different data sets
that often tend to draw the solution in different
directions. Moreover, MCE values of weighting parameters are objective, not dependent on users choice, although their values are usually close to those determined subjectively based on experience. Plotting various choices of weights typically results in
an L-shaped curve shown in Fig.~\ref{fig_Hertha_occ_weight}; the best
solution is at the corner of the curve. In this way, the
reconstruction from complementary data sources is possible even if no
single data mode is sufficient for modeling alone. 

For practical computations, the surface is rendered as a polyhedron, and $S$ and $h$ are computed accordingly with ray-tracing \citep{Kaa.ea:01}. Rather than using each vertex as a free parameter, the surface can be represented in a more compact form with spherical harmonics series (for starlike or octantoid; i.e., generalized starlike, shapes) or subdivision control points \citep{Kaa.Vii:12,Vii.ea:15}. These \emph{shape supports} are essential for convergence: they allow flexible modifications of the surface with a moderate number of parameters while not getting stuck in local minima  or over-emphasizing the role of regularization functions when searching for the best-fit solution. Each shape support has its own characteristic way of representing global and local features. For example, the octantoid parametrization
\begin{equation}
\label{eq:oct} \mathbf{x}(\theta,\varphi)=
\left\{\begin{array}{lr}
x(\theta,\varphi)=& e^{a(\theta,\varphi)}\sin\theta \cos\varphi,\\
			    y(\theta,\varphi)=&e^{a(\theta,\varphi)+b(\theta,\varphi)}\sin\theta \sin\varphi,\\
			    z(\theta,\varphi)=&e^{a(\theta,\varphi)+c(\theta,\varphi)}\cos\theta,
                             \end{array}\right.
\end{equation}
where $a$, $b$, and $c$ are linear combinations of the (real) spherical harmonic functions $Y^m_l(\theta,\varphi)$, with coefficients $\{a_{lm}\}$, $\{b_{lm}\}$, and $\{c_{lm}\}$, respectively, is easy to regularize globally while retaining the ability to produce local details.
The coordinates 
$(\theta,\varphi)$, $0\le\theta\le\pi$,  $0\le\varphi<2\pi$, 
parametrize the surface on the unit sphere $S^2$ but do not 
represent any physical directions such as polar coordinates.

\begin{figure}[t]
    \epsscale{1}
    \plotone{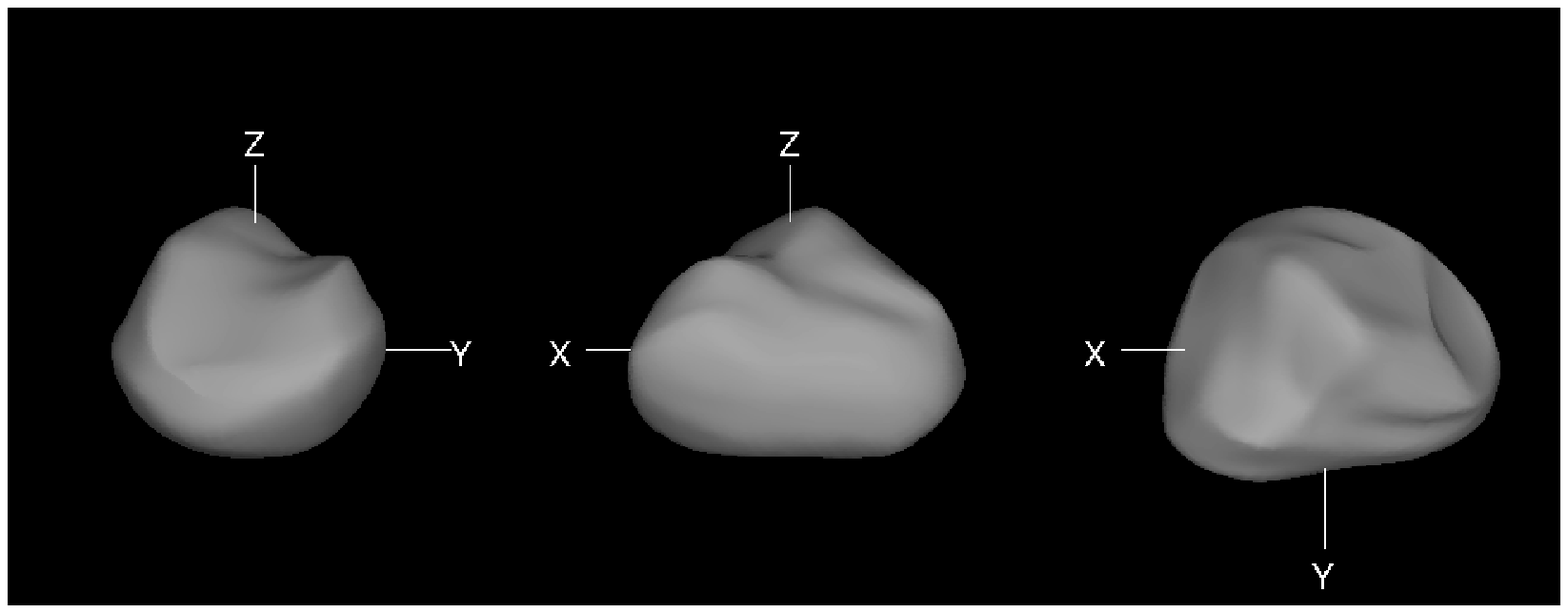}
    \plotone{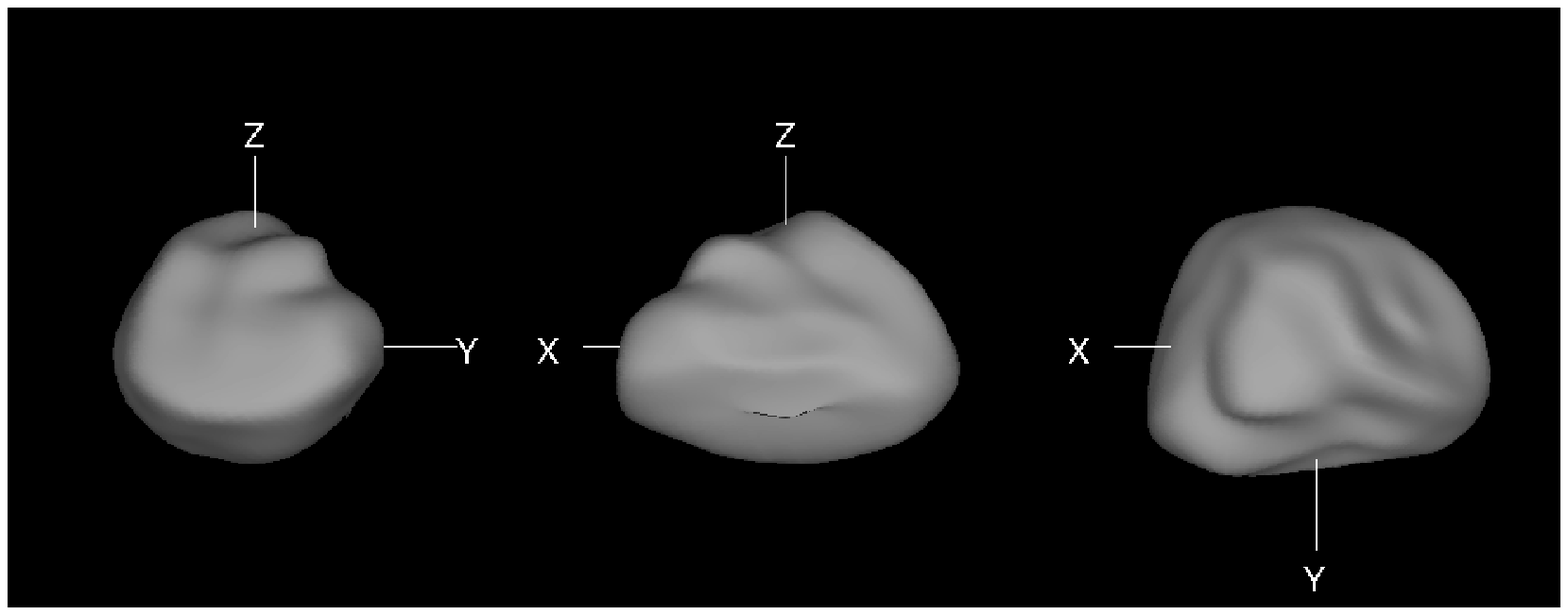}
    \caption{\small Model of (41)~Daphne reconstructed from
      lightcurves and adaptive-optics images using subdivision surfaces (top) and
      octantoids (bottom).
      The general shape remains stable, even if small-scale
        features slightly change.
    }  
    \label{fig:Daphne_shapes}
    \end{figure}

This inverse problem is a typical example of a case where model and
systematic errors dominate over random measurement errors. Thus the
stability and error estimation of the solution are best examined by
using different model types (Fig.~\ref{fig:Daphne_shapes}). In the case of shape, for example, the
reliability of the features on the solution can be checked by
comparing the results obtained with two or more shape supports
\citep[starlike, octantoid, subdivision;][]{Vii.ea:15}. This yields
better estimates than, e.g., Markov
  chain Monte Carlo sequences that only investigate
random error effects within a single model type. 

A particular feature of the model reconstruction from disk-resolved
data is that the result is dominated by the target image boundaries
rather than the pixel brightness distribution within the target
image. This is because the information is contained in the pixel
contrast which is the largest on the boundary (occultations are
special cases of this as they are samples of the boundary
contour). This is very advantageous when considering the effect of
model errors in luminosity properties (scattering or thermal models):
it is sufficient to have a reasonable model, and the result is not
sensitive to the parameters $L$. Thus, for example, Atacama Large Millimeter Array data can be
used for efficient reconstruction even with a very approximate
semianalytical Fourier-series thermal model -- more detailed models
have hardly any effect on the shape solution
\citep[Sect.~\ref{sec:interferometry},][]{Vii.Kaa:14,Vii.ea:15}. 

\section{\textbf{DATA AND MODELING}} \label{sec:data}
 
  We describe all data types that can be used, the way of
  collecting the data, their accuracy, typical number of asteroids for
  which the data exist, and expectations for the future. We also discuss a
  typical result of inversion -- what is the resolution of the model and
  how many targets can be modeled (Table~\ref{tab:tech}).
 
\subsection{\textbf{Photometry}}
\label{sec:photometry}
 
  Disk-integrated photometry is, and will always
  be, the most abundant
  source of data, because it is available for essentially every
  single known
  asteroid. Because asteroid brightness periodically changes with its
  rotation, frequency analysis of asteroid
  lightcurves provides asteroid rotation periods -- the basic physical property
  derivable from time-resolved photometry. The regularly updated
  Asteroid Lightcurve Database of \cite{War.ea:09} available at 
    \url{http://www.minorplanet.info/lightcurvedatabase.html}
  now contains rotation periods and other physical parameters for
  almost 7000 objects, for about half of them their rotation period is secure and unambiguous. The role of amateur astronomers in this field
  is traditionally strong, getting even stronger with increasing level
  of their technical and software equipment. Hundreds of asteroid lightcurves are published quarterly in the Minor Planet Bulletin; most of them are then archived in the Asteroid Light Curve Database at the Minor Planet Center site (\url{http://mpc.cfa.harvard.edu//light_curve2/light_curve.php}) in the ALCDEF standard \citep{War.ea:11}. The efficiency of lightcurve production can be increased by dedicated wide-field photometric surveys \citep[][for example]{Mas.ea:09, Pol.ea:12}, although the period determination from undersampled lightcurves is often ambiguous \citep{Har.ea:12}.

  For period determination, a single lightcurve covering the full
  rotation is sufficient. However, a set of such lightcurves observed
  at different geometries (asteroid illuminated and seen from various
  directions) is needed to reconstruct the shape and spin state of an
  asteroid. The {\em lightcurve inversion} method of \cite{Kaa.Tor:01,
    Kaa.ea:01} was already reviewed in {\em Asteroids III}
  \citep{Kaa.ea:02c}. Since then, the method has been widely used and
  hundreds of asteroid models have been derived. They are publicly available at the Database of Asteroid Models from Inversion Techniques (DAMIT, \url{http://astro.troja.mff.cuni.cz/projects/asteroids3D}).  The reliability of
  the method was proved by comparing its results with independent data
  such as laboratory asteroid model \citep{KaaS.ea:05}, 
  adaptive-optics images \citep{Mar.ea:06},
  stellar occultations \citep{Dur.ea:11}, or
  spacecraft images of asteroids
  (2867)~\v{S}teins \citep{Kel.ea:10} and (433)~Eros \citep{Kaa.ea:02c}.  
  
  From disk-integrated photometry alone, only a global shape without
  any small-scale details can be derived. Because the reflectivity of the surface is not known, the models are not scaled and the information about the size has to come from complementary data. To avoid over-interpretation
  and artifacts of the modeling, the shapes are usually represented by
  a \emph{convex} model.
  This allows to work not in the obvious
  radius parameter space but in the Gaussian image space (describing a
  convex body by the curvature of its surface). 
  This is less intuitive
  but it makes the inverse problem less vulnerable to errors of data
  and model because of the Minkowski stability \citep{Lam.Kaa:01} --
  even if the areas of individual surface facets may change
  significantly for slightly different data sets,  
  the global convex shape changes very little.

  From the practical point of view, we are interested in
  finding a {\em unique solution} of the inverse problem. To guarantee
  this, observations covering a sufficiently wide range of
  viewing and illumination geometries are needed. For a typical main-belt
  asteroid, it means observations during several apparitions. For a
  close-approaching near-Earth asteroid, several months could be
  sufficient. For more distant objects, we can in principle observe changing lightcurve amplitude because of changing aspect, this would anyway take many decades. 
  Nevertheless, trans-Neptunian objects (TNOs) and Centaurs can be never observed
  from Earth at geometry significantly different from opposition, which is not
  sufficient to reconstruct a unique convex model \citep{Rus:1906}. If the
  lightcurves observed in various filters are different, there is some
  spectral/color variegation over the surface, and 
  a crude color map can be reconstructed \citep{Nat.ea:05}.  

\begin{figure}[t]
 \plotone{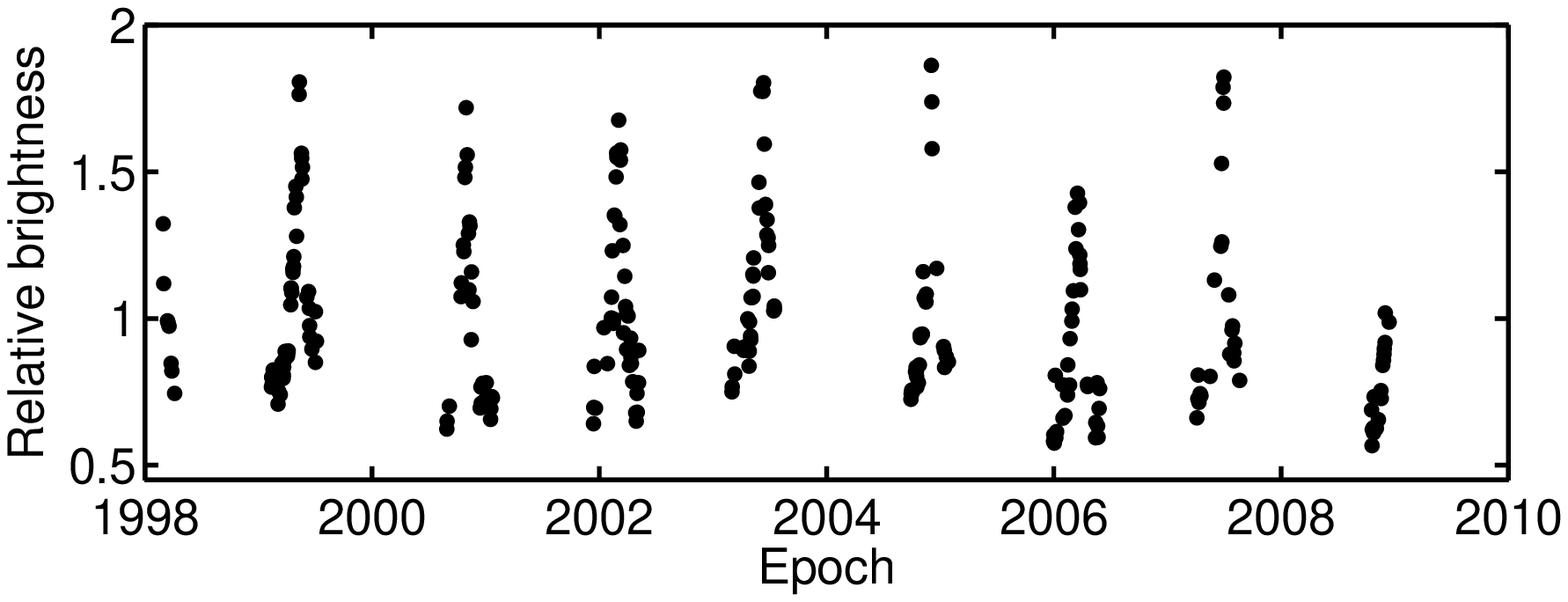}\\
 \plotone{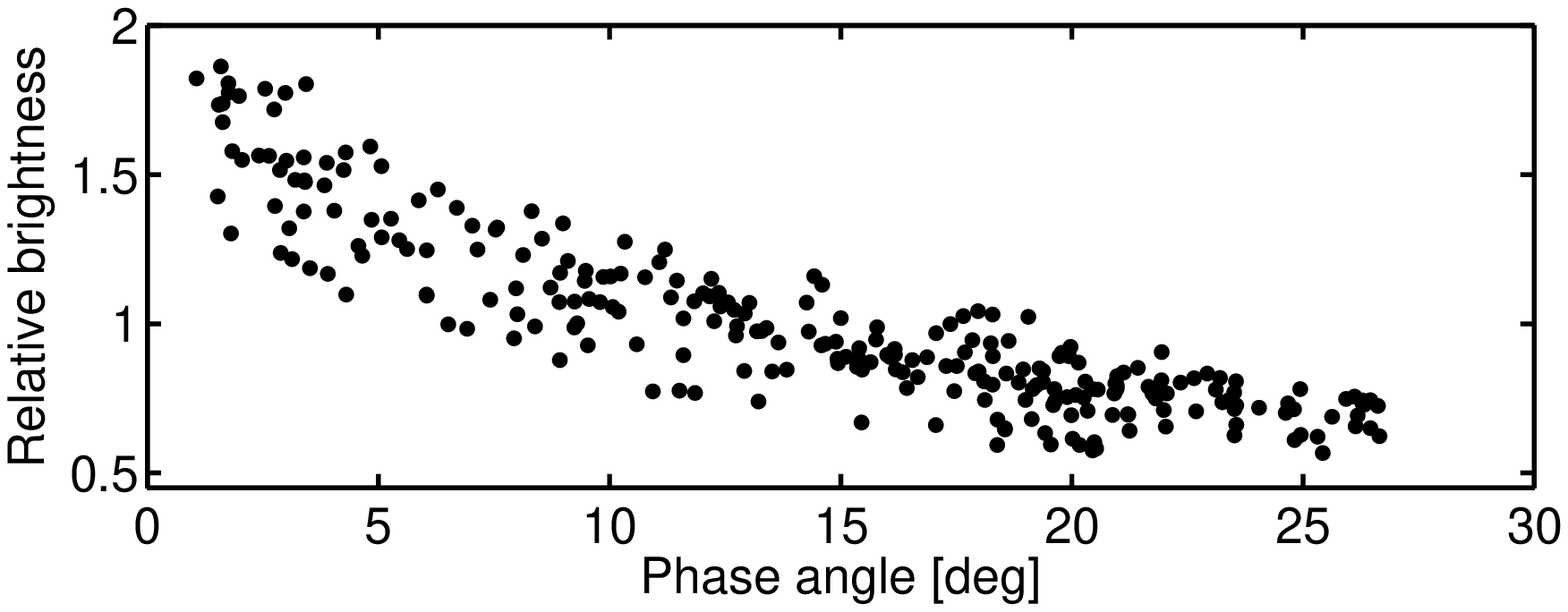}\\
 \plotone{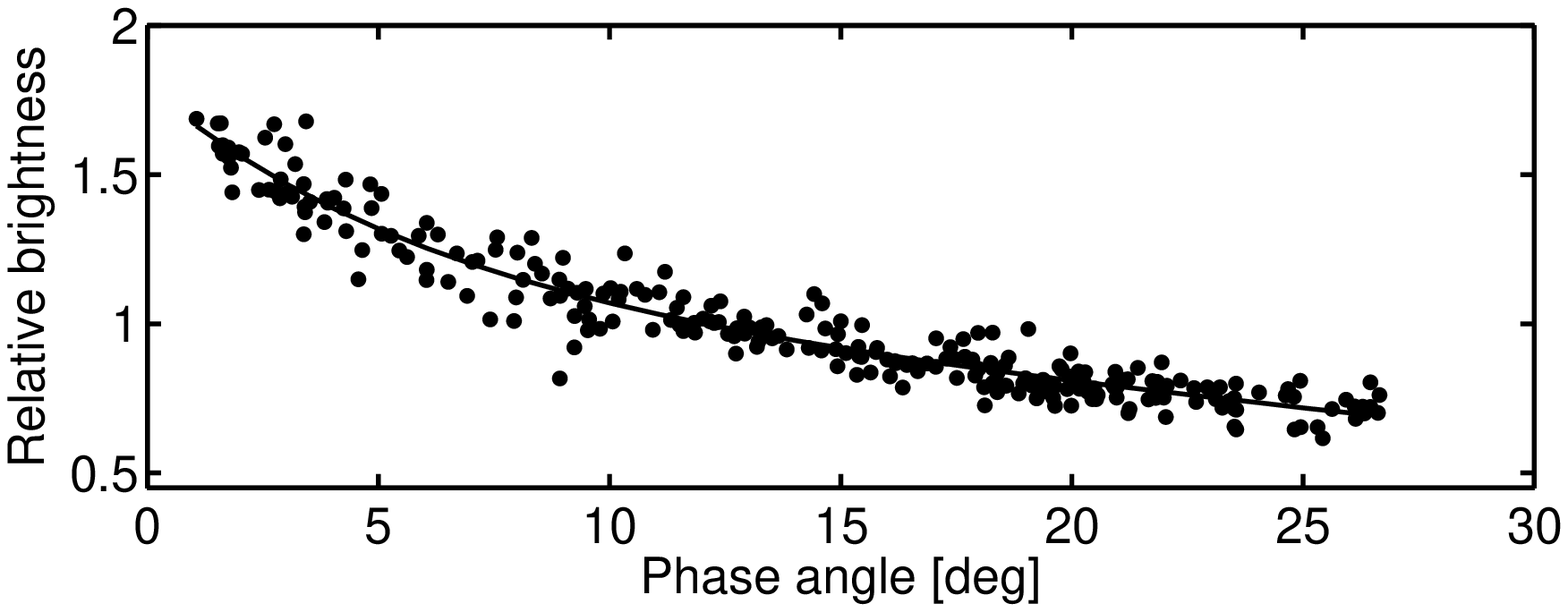}\\
 \caption{\small Sparse-in-time photometry of (21)~Lutetia obtained at
   US Naval Observatory plotted as a function of time (top) and phase
   angle (middle). The brightness was reduced to the unit distance
   from Earth and Sun. The scatter in the phase plot is caused
   by the intrinsic noise of the data and the rotational and aspect
   effects. The bottom panel shows residuals for the best-fit model
   (Fig.~\ref{fig_Lutetia_models}) plotted over the phase curve (solid curve).
 }  
 \label{fig_Lutetia_sparse_LC_fit}
\end{figure}

A typical outcome of lightcurve inversion is a convex shape model that describes the global characteristics of the real
asteroid. It is also easy to obtain nonconvex versions with the general inversion procedures (see Sect.~\ref{sec:inversion}), but then one
should produce several solutions with various model types and parameters, and be very cautious about the results \citep{Vii.ea:15}. 
In general, disk-integrated photometry contains very little information about nonconvexities unless they are
very pronounced or observed at very high phase angles where shadowing effects play an important role \citep{Dur.Kaa:03}.
Nonconvex models seldom fit lightcurves better than convex ones simply because the latter typically already fit the data down
to noise level. This sets the resolution limit of photometry \citep[see the discussion in][and references therein]{Kaa.ea:01, Kaa.ea:02c,
Vii.ea:15}.

The relative accuracy of the sidereal rotation period determined from
lightcurves is of the order of $10^{-5}$ or better, depending mainly
on the time span of observations. The direction of the spin axis can
be determined with an accuracy from a couple of degrees for the models
based on many decades of observations with many dense lightcurves, to
more than twenty degrees for models based on limited and noisy
data. For asteroids orbiting in the ecliptic plane, the geometry for
an Earth-based observer is limited to that plane. Then the
disk-integrated brightness of a body with surface described by radius
vector $(x,y,z)$ and pole direction $(\lambda, \beta)$ in ecliptic
coordinates is the same as for a body $(x,y,-z)$ with the spin axis
direction $(\lambda + 180^\circ, \beta)$ \citep{Kaa.Lam:06}. That is
why for a typical main-belt asteroid, there are usually two equally
good mirror shape solutions with about the same pole latitude and pole
longitude difference of about 180$^\circ$. This ambiguity can be
removed with disk-resolved plane-of-sky projections
(e.g., images, stellar occultations, see Sect.~\ref{sec:img}, \ref{sec:occ}). 

\begin{figure}[t]
 \plotone{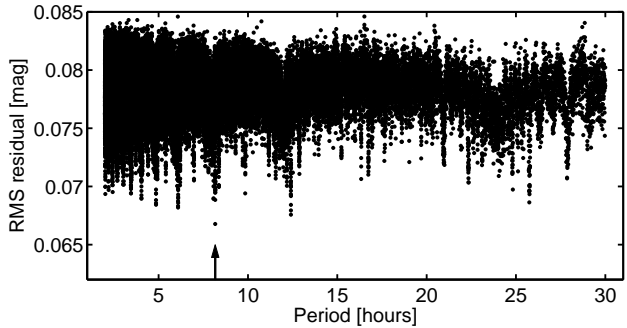}
 \caption{\small Period search for (21) Lutetia. Each point represents a local minimum in the parameter space to which the optimization algorithm converged. The lowest RMS residual (arrow) corresponds to the best-fit model shown in Fig.~\ref{fig_Lutetia_models}.}  
 \label{fig_Lutetia_period_search}
\end{figure}

  Apart from ``classical'' lightcurves where the sampling of brightness
  variations is dense with respect to the rotation period, there are
  also data that are {\em sparse in time}. Such data sets typically
  consist of only a single to a few brightness measurements per night.
  Provided the whole sparse data set is internally
  calibrated, it can be used the same way as a standard lightcurve
  that would be extremely long and very sparsely sampled \citep{Kaa:04}.
  These data are routinely provided by all-sky astrometric
  surveys with various -- usually poor -- photometric quality. Current
  surveys provide data of accuracy scarcely better than
  0.1\,mag. Given a typical lightcurve amplitude of 0.3\,mag, the
  signal is often drowned in noise and systematic errors. This leads
  to many physically acceptable models fitting the data to the noise
  level \citep{Dur.ea:05}. However, the total amount of data is huge
  and it has been shown that at least for some asteroids, models can be
  successfully derived from these data or from their combination with
  dense lightcurves \citep{Dur.ea:09, Han.ea:11}.  

  As an example, we show sparse photometry for asteroid (21)~Lutetia from US Naval Observatory (260 points covering ten years) in Fig.~\ref{fig_Lutetia_sparse_LC_fit} and the period search results in Fig.~\ref{fig_Lutetia_period_search}. The best-fit model based on this data set has the same rotation period and a similar spin axis direction as the models based on much larger and multiple data sets (Fig.~\ref{fig_Lutetia_models}).

  When using noisy sparse data or only a limited number of data points, a simple shape model of a triaxial ellipsoid is usually sufficient to model the data and to derive the correct period and spin vector orientation \citep{Cel.ea:09,Cel.Del:12,Car.ea:12b}. The advantage of this approach is that the shape is described by only two parameters (axes ratios) and the scanning of the period parameter space is much faster than with general shapes.

  The on-going (Pan-STARRS, Catalina, Gaia) and future (ATLAS,
  LSST) surveys will provide new data every night for
  essentially every known asteroid. Using this data for automated
  lightcurve inversion with well-mastered treatment of systematic
  effects, recognition of the best-fit models, definition of
  uniqueness of the solution etc., is the main challenge for the future
  lightcurve inversion.  
 
 \subsection{\textbf{Remote sensing disk-resolved images}\label{sec:img}}

  The most direct way to obtain information on the
  shape of an asteroid is to take pictures of it. The apparent shape
  as visible on the plane of the sky is delimited by the limb and the
  terminator, and multiple views obtained while the target rotates
  can fully characterize its 3-D shape.
  To resolve the small angular diameter ($<$\,0.5\arcsec)
  sustained by asteroids, large facilities are, however,
  required.

  In the 1980s, speckle imaging or speckle
    interferometry provided 
    ``\textit{the first glimpses of an asteroid's surface}''
    \citep[i.e., (4) Vesta by][]{1988Icar...73....1D}.
    This technique is based on  the analysis of the speckle pattern in
    the images of astronomical sources obtained through large telescopes at
    high magnification power and very short integration time. The aim is to
    overcome the blurring effects due to the  astronomical seeing and to
    attain diffraction-limited resolution images. Speckle
    interferometry has been commonly used to study the size, shapes, and surface features of the largest
    asteroids \citep{1985Icar...61..132D,1988Icar...73....1D,Rag.ea:00,Cel.ea:03}. 
    With the launch of the Hubble Space Telescope (HST) in orbit and the
    first light of the large (10-m class) ground-based telescopes equipped with
    adaptive-optics fed cameras (e.g., W. M. Keck, European
      Southern Observatory (ESO) Very Large Telescope (VLT), Geminis,
    Subaru), the importance of speckle interferometry has decreased.

  The critical issue in direct imaging is of course the angular
  resolution. Any image is the result of the convolution of the object
  on the plane of the sky with the instrument response, the point
  spread function (PSF).
  In space, the PSF is stable and corresponds to the diffraction
  pattern of the telescope. From the ground, the atmospheric
  turbulence constantly deforms the PSF and blurs the images, hence the
  need of real-time correction of the PSF by adaptive optics (AO).
  The technical challenges of sending a large telescope in space and
    of building deformable mirrors, explain why the first disk-resolved images
  in the 1990s were 
  still limited in resolution and only the largest asteroids
  (1)~Ceres, (2)~Pallas, and  
  (4)~Vesta have been imaged. In the decade since
  {\em Asteroids III\/}, numerous studies have been based on direct imaging of 
  asteroids
  \citep[e.g.,][]{2005-Nature-437-Thomas,
    2008-AA-478-Carry,
    2008-Icarus-196-Descamps, 
    2009-Science-326-Schmidt, 
    Mar.ea:13}.

\begin{figure}[t]
    \centering
    \epsscale{1}
    \plottwo{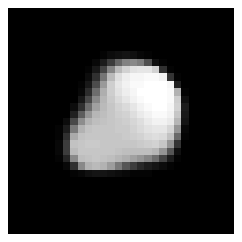}{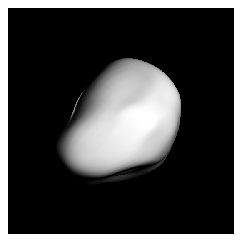}
    \caption{\small Deconvoluted adaptive optics image (left) of
      asteroid (41)~Daphne, and the corresponding image of the
      reconstructed model (right, projection of the the octantoid model from Fig.~\ref{fig:Daphne_shapes}). The model is shown under artificial illumination that enhances its 3-D shape and is different from the simple light-scattering model used for the inversion. The AO image has resolution of	
      10\,mas/pixel. The model was reconstructed from 14 AO images
      obtained with ESO VLT and several lightcurves.} 
    \label{fig:Daphne_AO}
    \end{figure}

\begin{figure}[t]
 \epsscale{1}
 \plotone{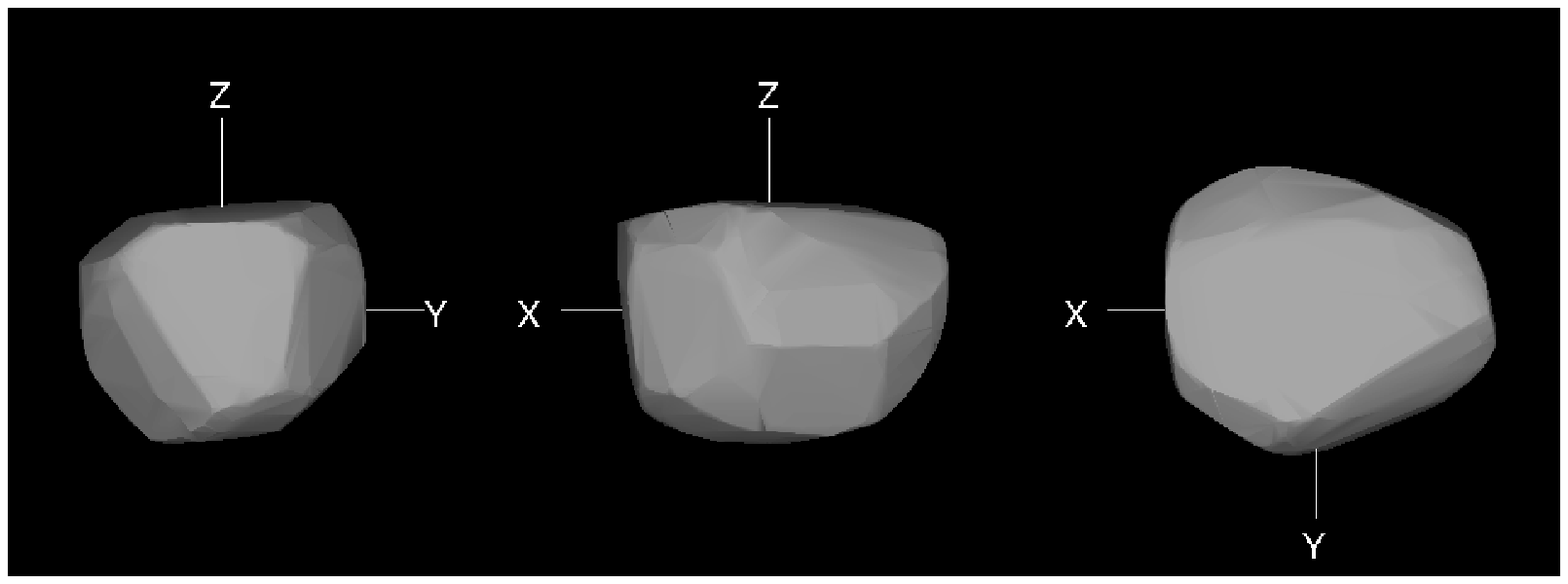}
 \plotone{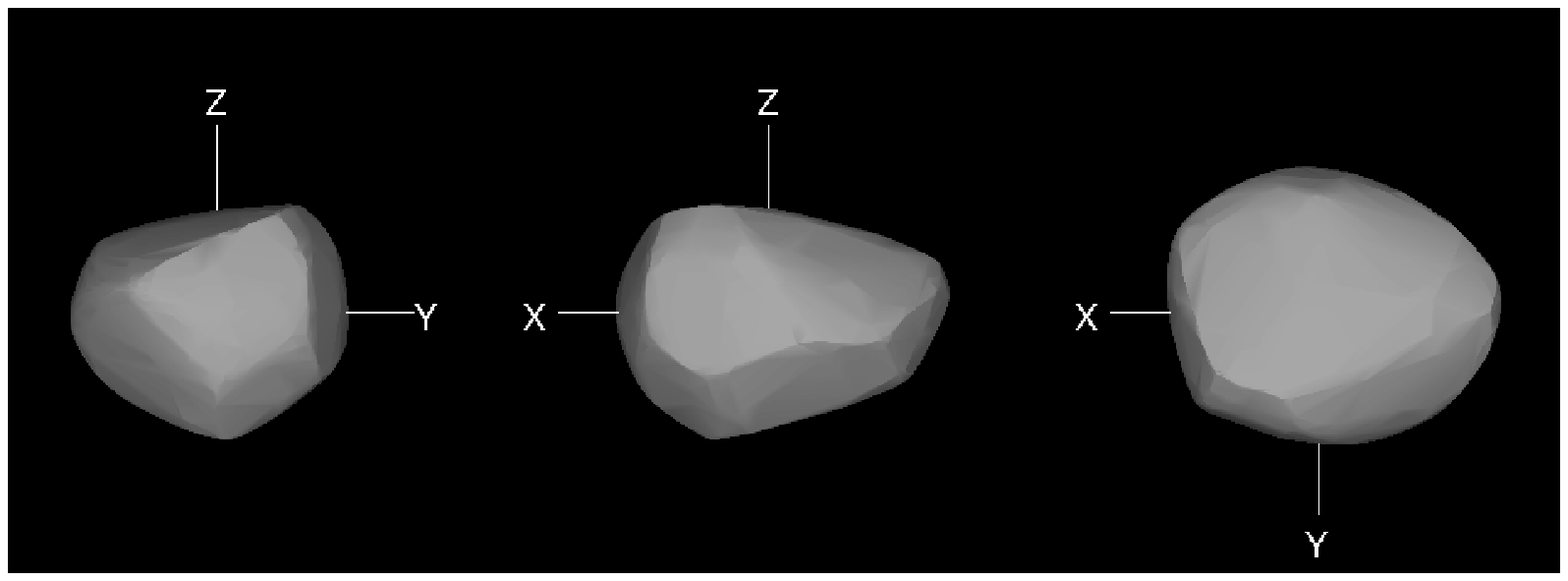}
 \plotone{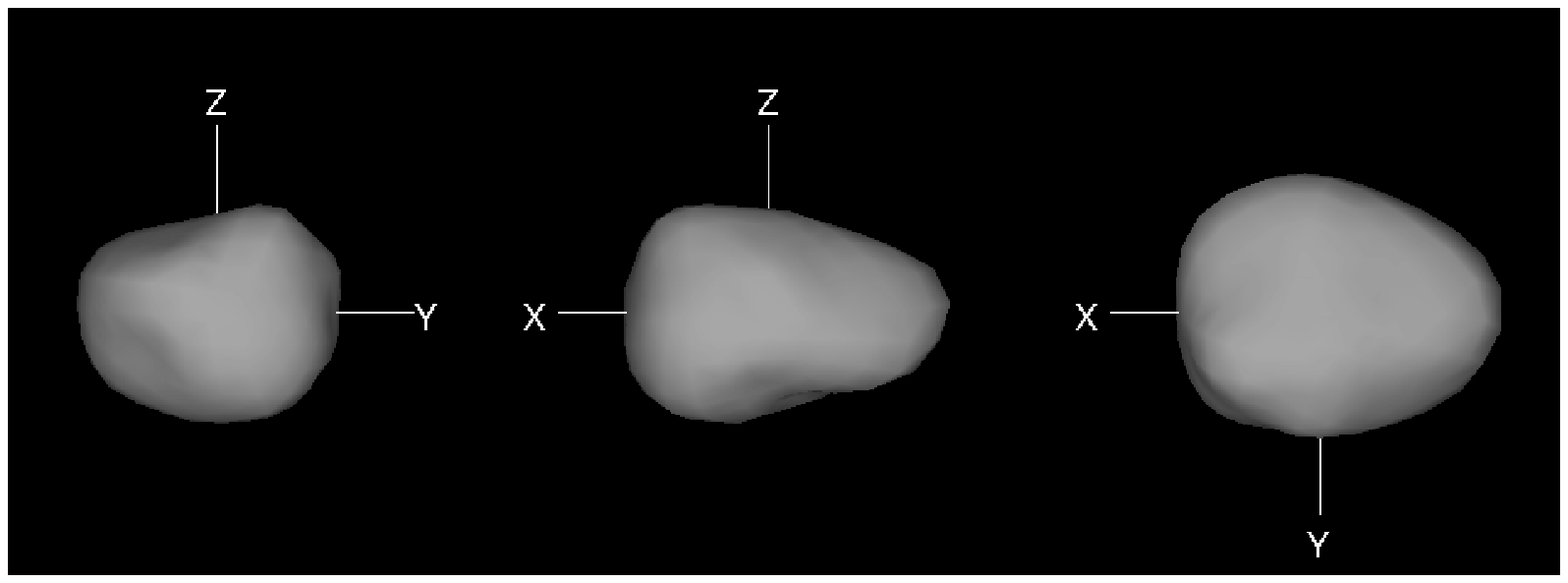}
 \plotone{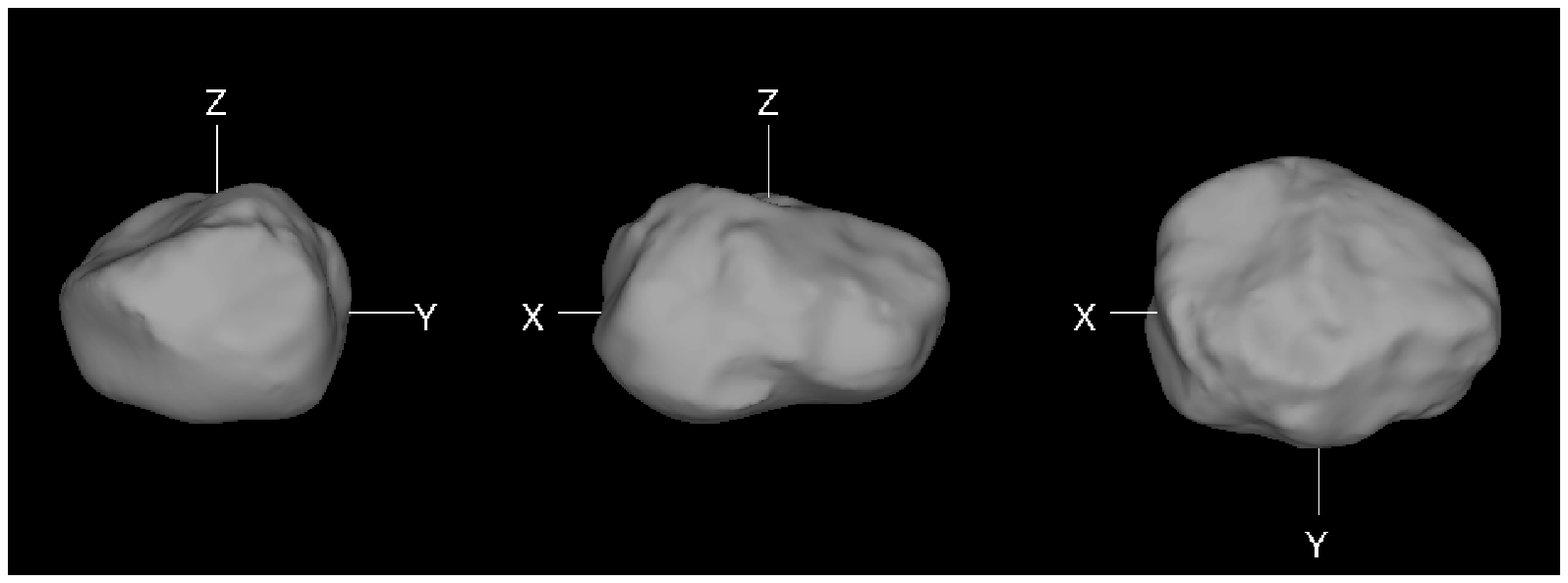}
 \caption{\small Different shape models for (21) Lutetia derived from
   different sets of data. The level of detail increases with more
   data from top to bottom: sparse photometry (Sect.~\ref{sec:photometry}),
   dense lightcurves \citep{Tor.ea:03},
   lightcurves and AO \citep{2010-AA-523-Carry},
   Rosetta flyby \citep{Sie.ea:11}.} 
 \label{fig_Lutetia_models}
\end{figure}

  Both from space or with adaptive optics on the ground,
  however, the contrast and angular resolution can be 
  improved by deconvolution of the image by the PSF. This is
  particularly true for images acquired from the ground with residuals
  from non-perfect AO correction.
  Deconvolution is an ill-posed problem,
  but robust algorithms adapted to planetary images are available
  \citep{2000-Msn-99-Conan, 2004-JOSAA-21-Mugnier, 2007-JOSAA-24-Hom}
  and have been validated on sky \citep{2006-JGR-111-Witasse}.
  It is nevertheless the most critical part of the post-processing, as
  an incorrect deconvolution can introduce a systematic error on the
  apparent size. 
  An example of an AO image and the reconstructed shape model for
  asteroid (41)~Daphne is shown in Fig.~\ref{fig:Daphne_AO} 
  
  Current facilities deliver an angular resolution of about 30 to
  50 milliarcseconds (mas),  
  depending on the wavelength. The apparent shape can therefore be
  measured for asteroids with an apparent diameter larger than 
  $\approx$\,80--100 mas, i.e., a couple of hundred targets.
  Simulations and observations of known targets such as the satellites
  of Saturn
  \citep{Mar.ea:06, 2009-DPS-41-Drummond,
    2009-PhD-Carry} have shown that a precision of a few mas can be
  derived on the 2-D profile on the plane of the sky, corresponding to
  only a few kilometers for main-belt asteroids.
  With upcoming large telescopes (30+\,m, such as the Thirty
    Meter Telescope or the European Extremely Large Telescope ),
  the angular resolution will
  be improved by a factor 3--4, providing more than 500
  targets. Second generation instruments with extreme AO foreseen on
  these telescopes should allow the observation of about 7000
  asteroids, with sizes of only a few kilometers 
  \citep{2013-Icarus-225-Merline}.
  
  Using disk-resolved images, 
  giant craters have been discovered
    \citep{1997-Science-277-Thomas, 2007-Icarus-191-Conrad},
  ambiguity in spin solutions have been solved
    \citep{Mar.ea:06, 2010-Icarus-205-Carry-a},
  albedo maps have been constructed
    \citep{2006-Icarus-182-Li, 2008-AA-478-Carry}, 
  convex 3-D shape models from lightcurves have been set to scale
    \citep{Han.ea:13a}, 
  and full 3-D shape models determined
    \citep{2010-AA-523-Carry, 2011-Icarus-211-Descamps}.

\subsection{\textbf{Flybys}}

  Asteroid imaging and mapping \emph{in situ} is essentially an
    extreme example of disk-resolved images.
  The modeling in such cases is more a
  cartographic than an inverse problem as the data are abundant,
  directly usable (containing identification points on the surface,
  etc.), and high-resolution (there is no ill-posedness). However,
  asteroid \emph{flybys}, during which roughly a half of the target
  is likely to remain unseen (not visible and illuminated), pose a special
  problem: how to see the dark side? The principle here is to use the
  high-resolution map of the seen side, constructed with a
  number of methods of photoclinometry, photogrammetry, and image fitting
  \citep{Pre.ea:12,Gas.ea:08,Jor.ea:12}, as a constraint in the
  otherwise same multi-mode inverse problem as with ground-based
  observations. The procedure is described in \cite{Kaa.Vii:12}; with
  it, the reconstruction of the dark side, such as those of 
  (2867)~{\v S}teins \citep{Kel.ea:10} or (21)~Lutetia
  \citep[][ see also Fig.~\ref{fig_Lutetia_models}]{Sie.ea:11}, is more detailed than from, e.g.,
  lightcurves alone. This is because half of the target is accurately
  reconstructed, with practically no error margin: therefore the
  fluctuation margin of the dark side is considerably smaller as
  well.

 \subsection{\textbf{Stellar occultations}\label{sec:occ}}

\begin{figure}[t]
 \epsscale{1}
 \plotone{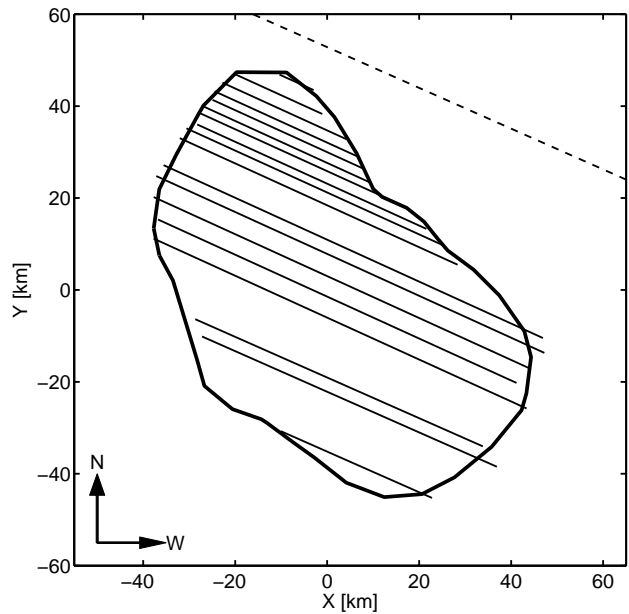}
 \caption{\small An example of a well-covered occultation that provides almost an ``image'' of the projected asteroid (135)~Hertha \citep{Tim.ea:09}. The model of Hertha based on its lightcurves and occultation data is over plotted. The dashed line is a negative observation constraining the northern part of the model's projection. Formal timing uncertainties are between 0.02 and 0.2\,s, corresponding to  0.12--1.2\,km in the projection, which is below the resolution of the model. The RMS of the fit is 1.9\,km.}  
 \label{fig_Hertha_occ}
\end{figure}

  The observation of a stellar occultation consists in recording the
  duration of the disappearance of a star behind the asteroid. Knowing
  the apparent motion of the asteroid on the plane of the sky,
  obtained from its ephemeris, this duration can be converted in a
  physical length on the disk of the asteroid, called a chord.
  Provided several observers record the same event from different
  locations on Earth, the 2-D profile of the asteroid is drawn on the
  plane of the sky \citep{1989-AsteroidsII-Millis}. 
  The main difference with disk-resolved imaging resides in the
  profile, made by only the limb for occultations, and limb+terminator
  for imaging. In fine, both techniques provide the 2-D profile of the
  target as projected on the plane of the sky at the epoch of
  observations (Fig.~\ref{fig_Hertha_occ}).

  Disk-resolved imaging and stellar occultation are, however,
  radically different in term of facilities, data processing,
  potential targets, reproducibility, and achievable precision.
  For stellar occultations, the properties of the occulted star matter
  generally more than the actual target: the asteroid.
  If the occulted star is bright enough, its occultation, even by
  a very small asteroid, can be detected with small aperture
  telescopes.
  This is of course the main advantage of stellar occultations, where
  the apparent size and shape of potentially any asteroid can be
  measured. Moreover, this technique can be successfully used also for distant TNOs, which have angular sizes too small to be resolved by imaging \citep{Sic.ea:11}. In practice, however, a given asteroid will only seldom
  occult bright stars. Measurements are thus hard, if not impossible,
  to reproduce.

  Stellar occultations are nevertheless extremely valuable. 
  The accuracy of the timing is dictated by
  time-series photometry, and can therefore be extremely precise.
  An uncertainty of 50\,ms in timings converts into only a 300\,m uncertainty
  in the length of the chord, typical for a main-belt asteroid (at
  1.5\,au from Earth with an apparent motion of 10\arcsec/h).
  The main source of uncertainty is, however, the absolute timing of
  each chord, required to align them on the plane of the sky.
  Most of historical occultations were recorded by naked eye,
  and suffered from this.
  Since a decade ago, thanks to the availability of low-cost positioning and timing
  systems (e.g., GPS), stellar occultations are being more and more
  valuable.

  The main contributors to the field 	
  are currently amateur astronomers: for a given
  event, observers have to move to set themselves on the
  predicted occultation
  path on Earth. Small aperture ($<$\,20\,cm) mobile stations are
  therefore ideal for recording stellar occultations.
  Because of the uncertainties of the star and asteroid positions on sky, there
  is generally an uncertainty of a few tens of kilometers in the
  location of the occultation path on Earth, requiring observers to
  spread over large area to cover the event. This usually prohibits dense
  coverage of the asteroid profile.
  When only a couple of chords are available, the event 
  provides only limited information on the size, if any on the shape. 
  Current occultation predictions of sufficient accuracy concern only stars in the Hipparcos catalogue and large (at least tens of kilometers) asteroids.
  With the upcoming publication of the ESA Gaia stellar catalog, and
  update of asteroid orbits, this position uncertainty is expected to drop
  significantly and future occultations will be easier to predict,
  hence observe \citep{2007-AA-474-Tanga}.

  Overall, stellar occultations can provide precise measurements of
  the size and shape of an asteroid, as projected on the plane of the
  sky. However, events are rare for a given target. Occultations are
  therefore very valuable in combined data sets, as for instance, to
  set scale to otherwise dimensionless 3-D shape models
  \citep[e.g.,][]{Dur.ea:11, Han.ea:11}.
  From almost 2500 occultations compiled by \cite{Dun.ea:14}, there are about 160 ``good'' ones that allow a reliable determination of asteroid's size and about 40 ``excellent'' ones that show 
  details in asteroid's profile. 

\subsection{\textbf{Interferometry}}
\label{sec:interferometry}
 
Another technique to overcome the limitations of small angular sizes of asteroids, in order to measure their sizes, shapes,
and possible presence of satellites, is interferometry.
An astronomical interferometer combines coherently (i.e., conserving the phase information) the light from two or more
apertures of the same telescope or of distinct telescopes spaced by a distance $B$. The spatial resolution (in radians) is of
the order of $\lambda/B$ where $\lambda$ is the wavelength. 

In the following, we give a basic introduction to interferometry of
asteroids. Further details can be found in the following works:
\cite{2010SerAJ.181....1J} and \cite{2013Icar..226..419M} for a broad
introduction to astronomical optical interferometry, 
methods and instrumentation; \cite{2009ApJ...694.1228D} for a description of the techniques, and the models for
deriving the size and basic shape proprieties of asteroids from the ESO Very Large Telescope Interferometer (VLTI) MID-infrared Interferometric instrument (MIDI) data; \cite{Carry:2014tf} for an extension of the technique of
\cite{2009ApJ...694.1228D} to the determination of the sizes and the
separation of binary asteroids; \cite{2011Icar..215...47M} for a
description and the extension of a thermophysical model to
the analysis of interferometric data of asteroids with the
aim of obtaining surface properties such as the thermal inertia. 

\begin{figure}[t]
 \plotone{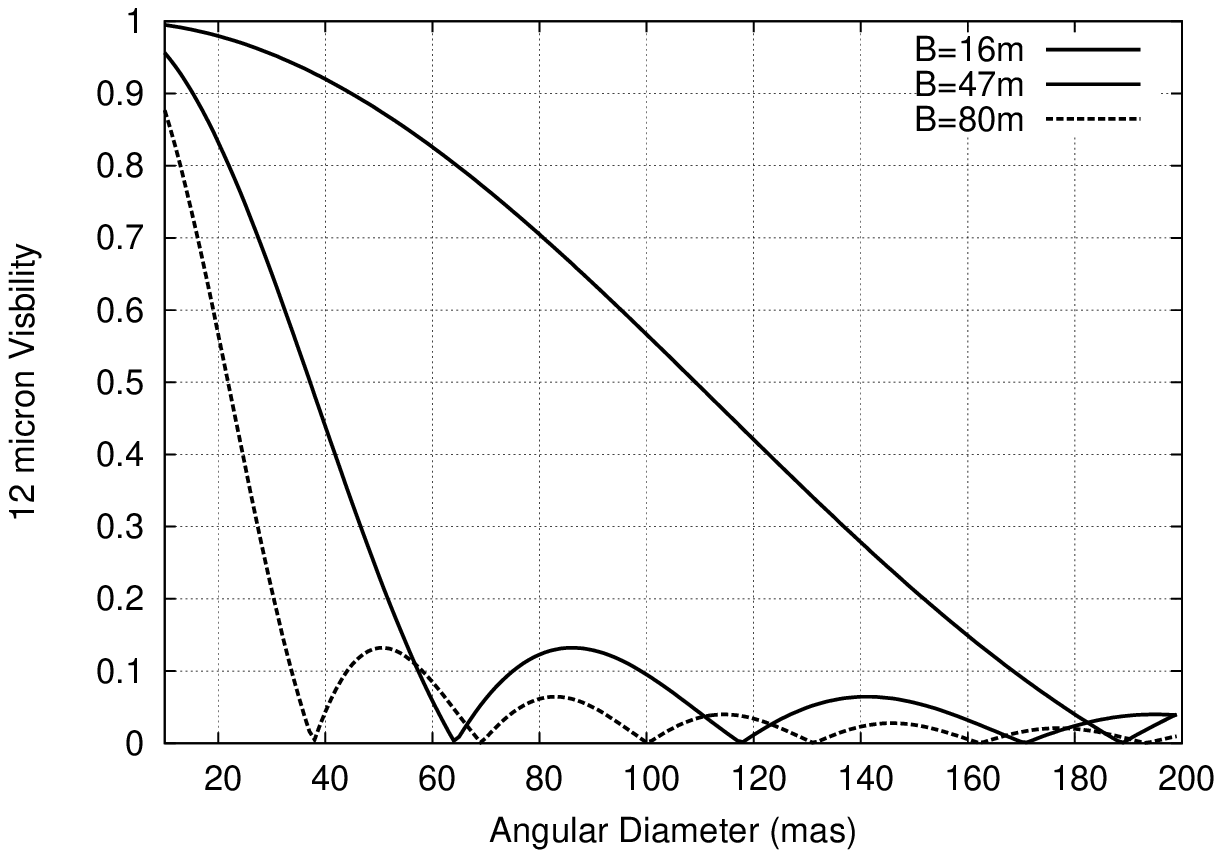}\\
 \plotone{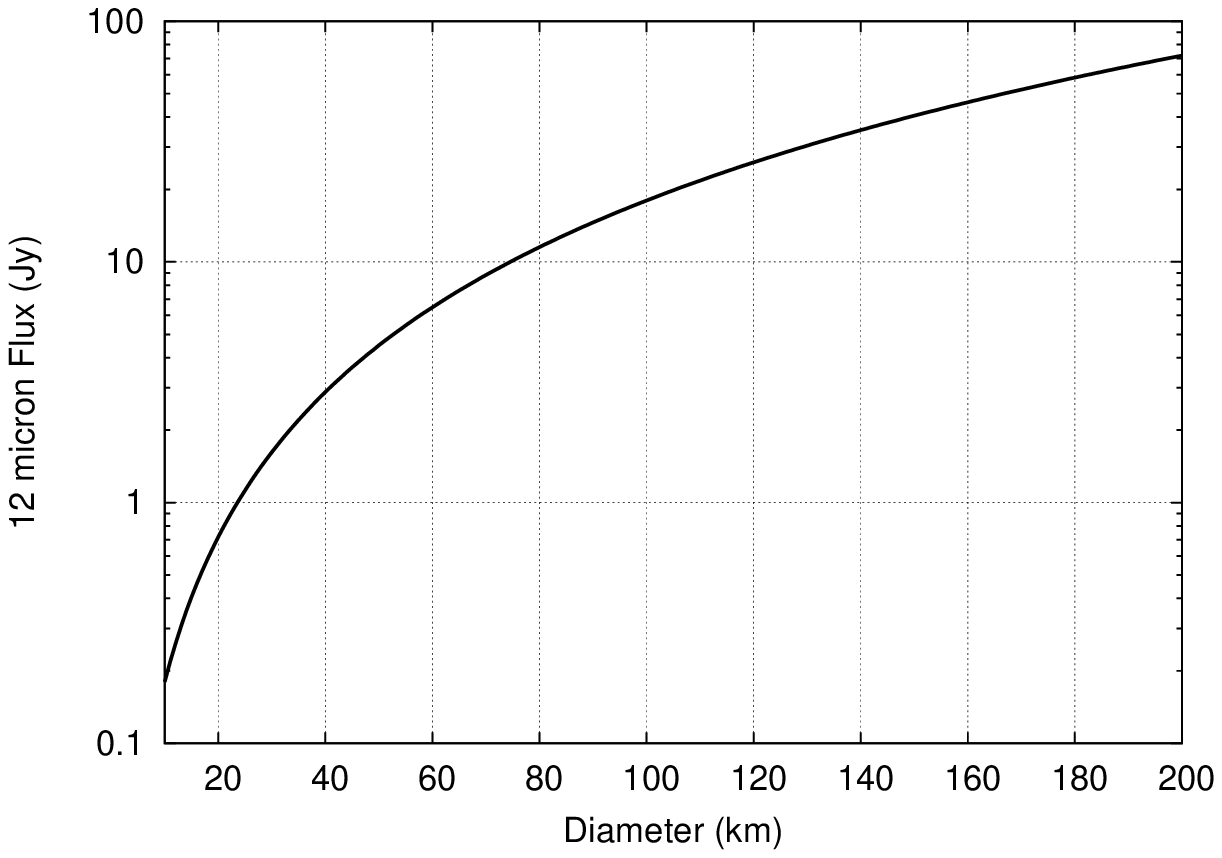}
 \caption{Visibility and total flux for an asteroid at 2.5\,au from 
  Sun and 1.5\,au from Earth as a function of its diameter for
  different values of the baseline. The MIDI limiting correlated flux (flux times visibility) is 0.5--1\,Jy and 10--20\,Jy for
  the UTs and the ATs respectively. Similar limiting fluxes are expected for MATISSE. See text for further details.}
\label{F:interf}
\end{figure}

Interferometers measure the coherence function of the source, also called the interferometric visibility, which is given by
the ratio between the correlated and the total flux. The correlated flux is the amount of flux in the interferometric fringes.
More precisely the correlated flux is the Fourier transform of the brightness distribution of the source measured on the
interferometer's baseline projected on the plane of the sky \citep[see][for example, and references
therein]{2009ApJ...694.1228D}. 

The Fine Guidance Sensors (FGS) aboard the Hubble Space Telescope (HST) are optical white-light shearing interferometers
\citep{1998SPIE.3350..237N} that combine the light from distinct apertures of the HST primary mirror and have been used to
measure the size and basic shape properties of asteroids \citep{2002A&A...391.1123H,2003A&A...401..733T}. HST/FGS data have
clearly demonstrated the bi-lobed nature of some bodies such
as (216)~Kleopatra \citep{2001Icar..153..451T} and (624)~Hektor \citep{2003A&A...401..733T}, but were not able to detect the
presence of the little moons orbiting these asteroids \citep{2011-Icarus-211-Descamps,Mar.ea:14}, due to the large magnitude
difference between the asteroids and their respective moons. Despite the impressive angular resolution of a few mas,
corresponding to a few kilometers at a distance of 1.2--1.5\,au, a clear limitation of  HST/FGS asteroid studies is the
bright limiting magnitude of the instrument of about $V\,\sim\,13$--14\,mag \citep{2003A&A...401..733T}. A recapitulation of the
HST/FGS asteroids results 
can be found in \cite{2002aste.conf..219D}. 

Ground-based interferometry is limited to observation of bright targets because of the requirement to take exposures shorter
than the atmospheric coherence time of a few milliseconds. The limiting magnitudes of ground-based long-baseline
interferometers and the intrinsic low surface brightness of asteroids, have prevented the use of these instruments for the
studies of small solar system bodies until the availability of the VLTI.  

Interferometry in the mid-infrared was proven to be also very sensitive to the global shape of asteroids and also to their
surface characteristics \citep{2011Icar..215...47M,2013Icar..226..419M}. In particular, since observations are typically
carried out in the thermal infrared (8--13\,\micron), MIDI data are sensitive to the surface temperature distribution, which
is strongly affected by the value of the thermal inertia ({\it Delbo et al.}, this volume).

Figure~\ref{F:interf} shows the visibility and the total flux as a function of the angular diameter of an asteroid at 2.5\,au
from Sun
and at 1.5\,au from Earth, where 1\,mas roughly corresponds to 1\,km on the asteroid. Note that main-belt asteroids smaller
than
$\sim$\,20\,km cannot be easily observed with MIDI at the VLTI.  
There are about thousand known asteroids with diameter above\,20 km, implying that interferometry is potentially an
interesting technique for shape modelling. On the other hand, as interferometry in the thermal infrared is sensitive to
the spatial distribution of the temperature on the asteroid surface, this technique can be used to determine thermophysical
properties of asteroids when the body shapes are known as demonstrated by \cite{2011Icar..215...47M,2013Icar..226..419M}.

Each VLTI baseline can be used with MIDI only one at the time. MIDI will likely be decommissioned in the near
future to be substituted after 2016 by the Multi AperTure mid-Infrared SpectroScopic Experiment (MATISSE). This instrument
will combine up to four Unit Telescopes (UTs) or  Auxiliary Telescopes (ATs), allowing six simultaneous baselines. This feature will enable us to measure the spatial
distribution of the infrared flux along different directions. MATISSE will also measure closure phase relations and thus offer
an efficient capability for image reconstruction. In addition to the N band, the MATISSE will also operate in the L and M
bands. Unfortunately, MATISSE is not expected to be more sensitive than MIDI. 

Another second generation instrument at the VLTI, GRAVITY, that will combine the light from all four UTs, will offer further
improvements in spatial resolution compared to MIDI and MATISSE (though with more a more stringent $V\lesssim 11$\,limiting
magnitude). It will provide near-infrared adaptive-optics assisted precision narrow-angle (about 4\arcsec) astrometry at the
10\,$\mu$as level in the K band (2.2\,\micron). Both the reference star and the science object have to lie within
the $\sim$\,4\arcsec\ field of view. In imaging mode, GRAVITY can achieve a resolution of $\sim$\,3 mas in the near-IR
\citep{2008SPIE.7013E..69E}. The imaging mode can be interesting to precisely measure the sizes and the orbits of the
satellites of large asteroids, the latter with $V\lesssim 11$\,mag. 

AMBER is the current near-infrared focal instrument of the VLTI. It operates in the J, H, and, K bands (i.e., from 1.0 to
2.4\,$\mu$m).  
The AMBER limiting magnitude for asteroid observations is $V\sim 9$\,mag. There is only a handful of asteroids brighter than
this limit. These bodies are also the largest ones with angular extensions generally $>$100\,mas, implying a very low
visibility in the J-H-K with the UTs baselines. Although the photometric flux of these few asteroids is such that
$H<7.5$\,mag, their correlated magnitudes
due to the low visibilities are much higher than 7.5\,mag, preventing their fringe detection and tracking in the near-IR.

However, one of the most interesting instruments for asteroid ground based interferometry is the Large Binocular
Telescope Interferometer (LBTI). It consists of two 8.4-m telescopes mounted side by side in a single mount, with a 14.4-m
center-to-center spacing. This configuration offers a unique capability for interferometry of a Fizeau beam combination. This
offers a wide field of view ($\sim$10--20\arcsec) and low thermal background. For example, the LINC-NIRVANA instrument can --
in principle -- be able to resolve binary asteroids whose components are separated $>20$\,mas. This will allow splitting many
binary asteroids. 

\begin{figure}[t]
\centering
\includegraphics[width=0.32\columnwidth]{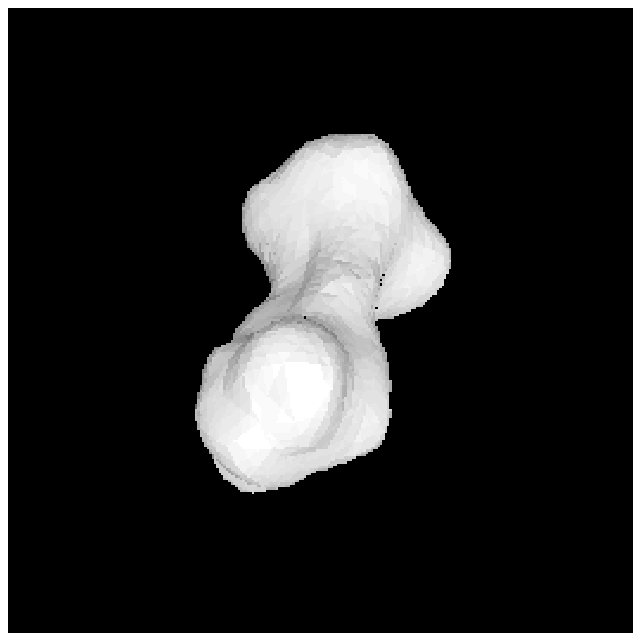}
\includegraphics[width=0.32\columnwidth]{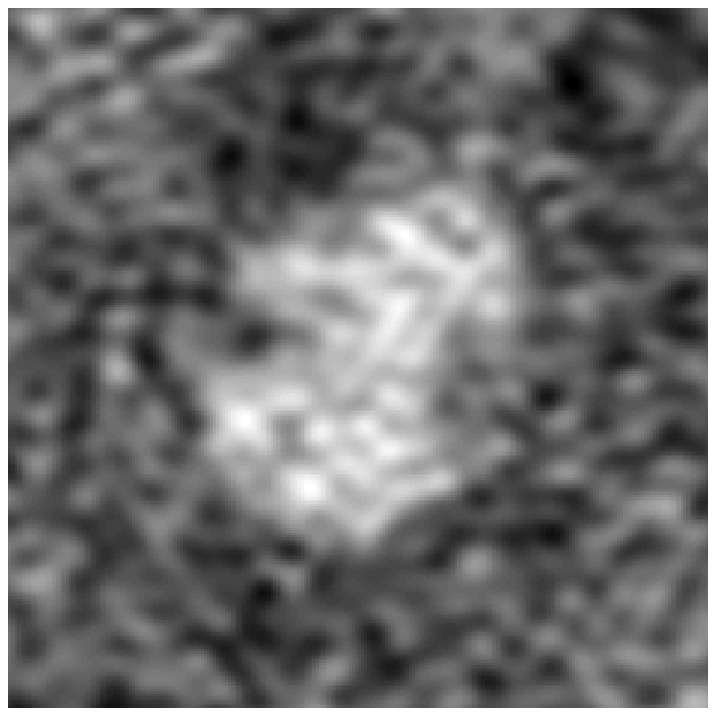}
\includegraphics[width=0.32\columnwidth]{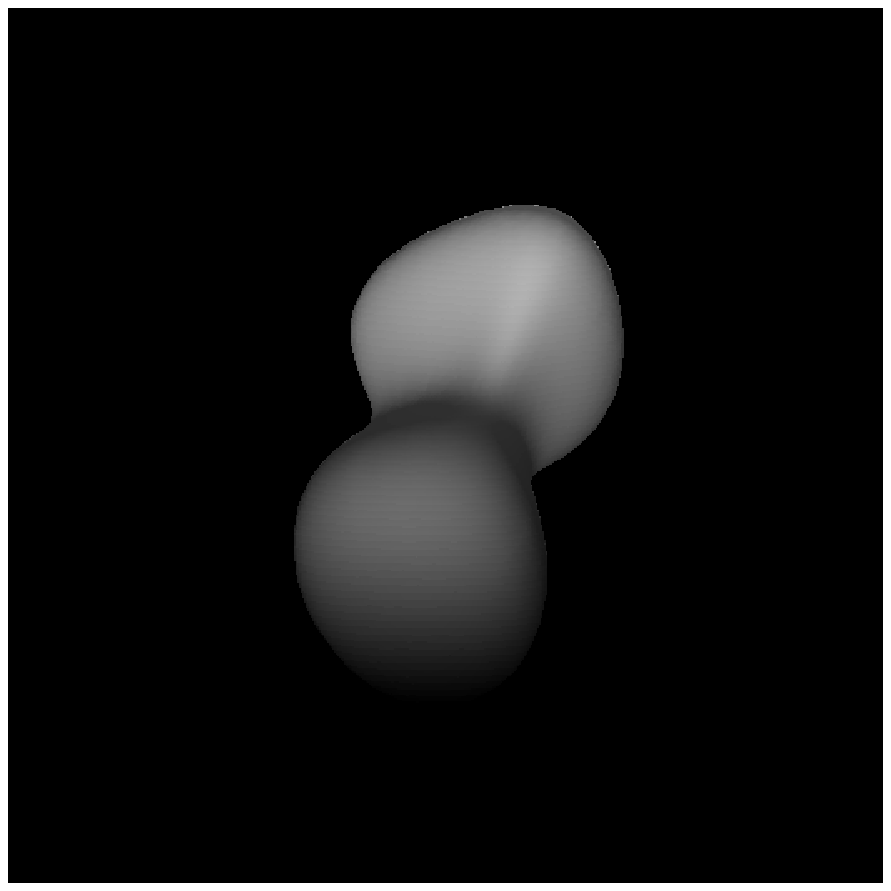}
\includegraphics[width=\columnwidth]{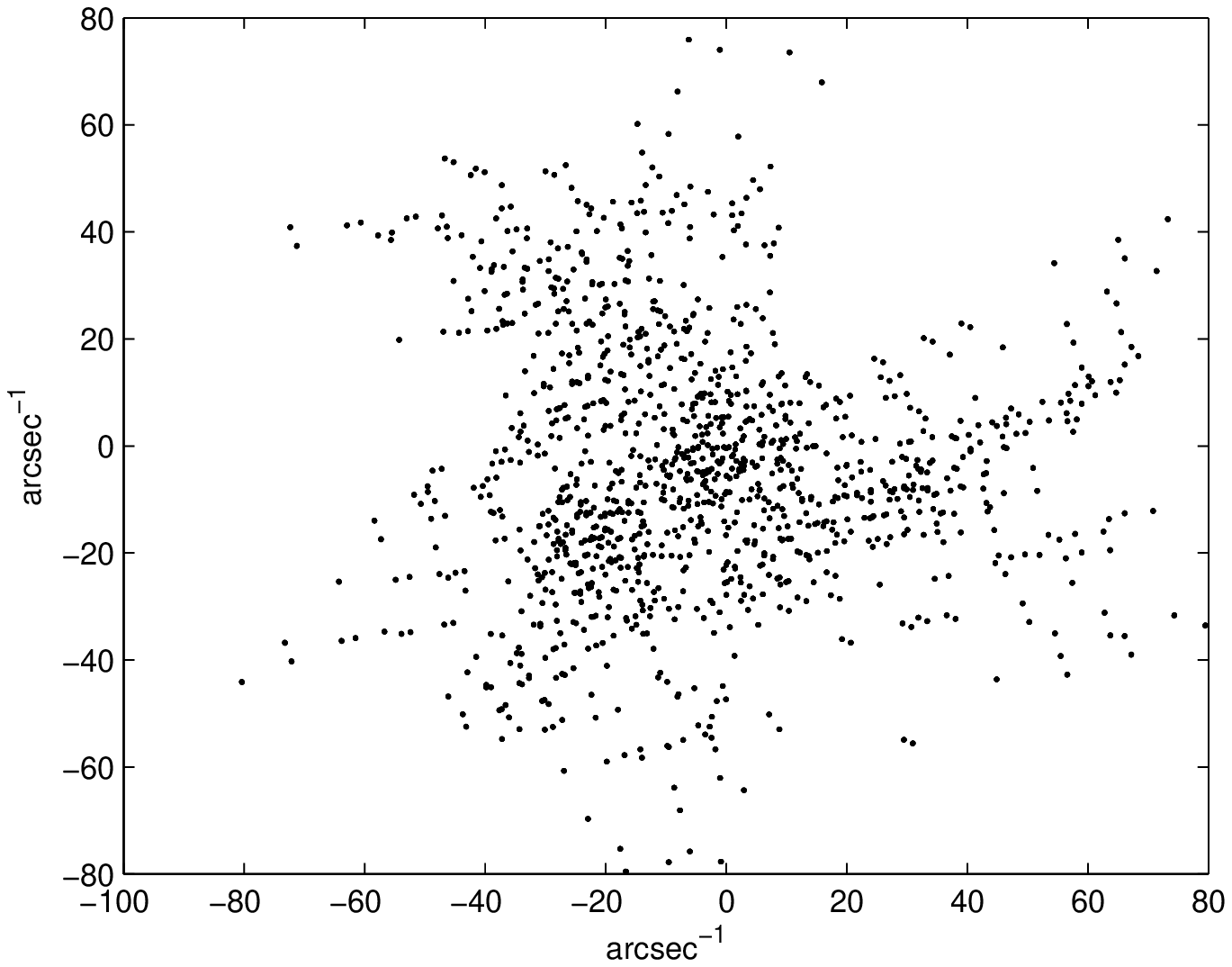} 
\caption{\small Simulated infrared flux from an asteroid  \citep[left, a radar shape model of Kleopatra by][]{Ost.ea:00} and a ``dirty image''
(center) with atmospheric noise, obtained by transforming the
incompletely sampled frequency plane (bottom, 6000 points, 50
antennas). The smallest resolvable detail is approximately
$10$\,mas.  Provided enough observations at different geometries are
available (here eight observation runs were used), a shape model (right) can be
constructed directly from the raw interferometric data.}
\label{fig_ALMA}
\end{figure}

Another important source of interferometric data in the near future is the Atacama Large Millimeter Array (ALMA). It will
provide
resolution $\sim$\,5\,mas at 0.3\,mm and a dense mesh of baselines, thus enabling ``imaging'' of hundreds of asteroids in the
main belt
\citep{Bus:09}.   
A useful feature of the multimodal inversion is that the raw ALMA (and
any other interferometric) data can be used directly as the original
Fourier transform: there is no need to
reconstruct the image 
estimate. What is more, the reconstruction of the overall shape is insensitive to
inaccuracies and uncertainties in the thermal model used because the model is determined mainly by the boundary of the projection, the distribution of brightness inside the boundary is much less important
\citep{Vii.Kaa:14,Vii.ea:15}. An example of inversion of simulated
ALMA data is shown in Fig.~\ref{fig_ALMA} 
 
Long-baseline interferometry can also represent a novel approach to determine the masses and the densities of asteroids in a
range of
sizes and distances never studied before. In particular, modern interferometers such as the VLTI, LBTI and the Magdalena Ridge
Observatory Interferometer can spatially resolve binary asteroids discovered by photometric lightcurves in the
main belt \citep{Carry:2014tf}. The separation of the components of these systems is too narrow for traditional observational
techniques
such as AO at 10-m class telescopes.

\subsection{\textbf{Direct size measurement with ESA Gaia}\label{sec:gaia}}

  The ESA Gaia mission, whose operations started in 2014, will provide
  accurate astrometry and photometry of asteroids. Photometric data can be used the same way as sparse photometry from ground-based observatories (Sect.~\ref{sec:photometry}). The potential of Gaia-like data was demonstrated by \cite{Cel.ea:09} on data from the Hipparcos satellite. Apart from disk-integrated brightness, Gaia will provide also  
  direct measurements of asteroid sizes, in a way much similar to the
  interferometry.
  Being designed as an astrometry mission at the $\mu$as level, 
  the PSF of Gaia is accurately known and stable. Deviations of the
  PSF from that of a point-like source can thus be measured and used
  to estimate the apparent size of asteroids.

  Owing to the amount of sources observed by Gaia, the satellite does not
  download images to Earth, but only 1-D flux profiles, corresponding
  to small 2-D windows centered on targets, stacked along one
  direction, similar to interferometry in that respect.
  Interpretation of this 1-D profile relies therefore on the
  \textsl{a priori}
  knowledge of the 2-D flux distribution on the plane of the sky.
  In the creation of Gaia catalog, this will be done iteratively,
  starting from simple spheres at zero phase angle to finally use
  the spin and tri-axial ellipsoid solutions determined
  otherwise from Gaia
  photometry \citep{2007-EMP-101-Mignard, 2007-AdSpR-40-Cellino}.

  These measurements represent a great opportunity for multi-data
  inversion algorithms: the measurement of the extension is direct,
  but clearly requires a realistic description of the projected shape
  on the plane of the sky.
  Considering Gaia specifications and observing geometry,
  \citet{2007-EMP-101-Mignard} have estimated the fraction of asteroids
  for which 1-D dimension will be measured with a precision better
  than 10\%.
  This fraction is highly dependent on diameter, and 20\% of asteroids
  between 20--30\,km will be measured at least once, while all
  asteroids larger than $\sim$\,80\,km will be measured repeatedly.

\subsection{\textbf{Disk-integrated radiometry}}

 Measurements of asteroids emission in the thermal infrared -- in
 general at wavelengths between 4--5 and 20--30\,\micron\ -- are
 mostly used to determine the sizes of these bodies \citep[{\it Mainzer et al.}, this volume; {\it Delbo et al.}, this
volume;][]{Delbo2002M&PS...37.1929D,Har.Lag:02}. Ground
 based telescopes can only observe in specific windows of the
 electromagnetic spectrum where the atmosphere is relatively
 transparent: i.e., the L, M, N, and Q bands at
   3, 5, 10, and 20\,\micron, respectively. However, such
 observations are  strongly affected by the variability of the
 transparency of the atmosphere and its thermal background. The
 background also receives contribution from the telescope and the
 optics. Hence, the thermal infrared observation of
 asteroids from the ground is limited to relatively bright asteroids
 ($V \lesssim 18$\,mag with 10-m class telescopes). Absolute
 calibration of the flux is
 rarely better than 5--10\%. On the other hand, in space the
 instrument calibration is usually stable and there is no need to
 reduce the thermal background from the atmosphere. As a consequence,
 space based telescopes such as Spitzer can observe much fainter and
 smaller asteroids \citep{Mommert2014ApJ...789L..22M}, with
 uncertainties in the calibration that can reach $\sim$\,1\% error. Also, from space the range of the observational
 wavelength is limited only by the detector technology, typically
 $\sim$\,3.5--50\,\micron. At longer wavelengths, the telescope optics
 require cooling and the observation of faint objects is confronted
 with the background from solar system dust cloud and infrared cirrus.  

In the \emph{Asteroids III} era, the main source of thermal
infrared observations of asteroids has been
 the IRAS Minor Planet
Survey that collected observations of more than 2200 asteroids
\citep{Tedesco2002AJ....123.1056T}. 
Since then, the NASA Wide-field Infrared Survey Explorer (WISE) has observed more
than 130,000 main-belt asteroids \citep[{\it Mainzer et al.}, this
  volume;][]{Masiero2012ApJ...759L...8M,Masiero2011ApJ...741...68M},
about 500 near-Earth asteroids
\citep{Mainzer2011ApJ...743..156M,Mainzer2012ApJ...752..110M,Mainzer2012ApJ...760L..12M},
about 1100 Hilda asteroids \citep{Grav2012ApJ...744..197G}, and almost
2000 Jupiter-Trojan asteroids
\citep{2012ApJ...759...49G,2011ApJ...742...40G} in four infrared
wavelengths at 3.4, 4.6, 12, and 22\,\micron;
the AKARI space telescope observed more than 5000 asteroids during its mission \citep{Usu.ea:11,Usu.ea:13,Has.ea:13};
the Spitzer space telescope observed hundreds of asteroids \citep[e.g.,][among others]{Eme.ea:06,Tri.ea:10,Lic.ea:12};
the Herschel Space Observatory, that, due to its longer wavelengths, spanning 55--671\,\micron, was primarily used to observe
trans-Neptunian objects \citep{Mul.ea:10}.
For a review about all these missions and their results see {\it Mainzer et al.}, this volume.

The thermal infrared spectrum of asteroids carries information about
their size and surface properties, such as the thermal
inertia, roughness and emissivity. 
These properties are typically derived by interpreting thermal
infrared data by means of thermal models \citep[{\it Mainzer et al.}, this
  volume; {\it Delbo et al.}, this volume;][]{Delbo2002M&PS...37.1929D,Har.Lag:02}. The
``simple'' thermal models that assume a
spherical shape, 
a Lambertian emission of the surface, and a simplified calculation of
the surface temperature distribution are used when we lack knowledge
of the asteroid global shape, spin vector and rotation period, which
is the majority of the cases. 
Widely used are, for example, the Near-Earth Asteroid Thermal Model
\citep[NEATM,][]{Harris1998Icar..131..291H,Delbo2002M&PS...37.1929D,Har.Lag:02}, 
or 
the Standard Thermal Model
\cite[STM,][]{Lebofsky1986Icar...68..239L,Har.Lag:02,Delbo2002M&PS...37.1929D}.

However, in order to derive the thermal inertia of an asteroid from
measurements of its thermal infrared emission, 
more sophisticated models, called thermophysical models (TPMs),
  are needed
 \citep{Spencer1990Icar...83...27S,Spencer1989Icar...78..337S,Lag:96a,Lagerros1997A&A...325.1226L,Lagerros1998A&A...332.1123L,Rozitis2011MNRAS.415.2042R,Mueller2007PhD,Delbo2004PhD}.
Such models are used to calculate the temperature distribution over the
body's surface as a function of different parameters, including the
thermal inertia. In these models, the asteroid shape is
usually fixed and is modeled as a mesh of planar facets. The
temperature of each facet is determined by numerically solving the
one-dimensional heat diffusion equation using assumed values of the
thermal inertia, with the boundary condition given by the
time-dependent solar energy absorbed at the surface of the facet (see
{\it Delbo et al.}, this volume, for a review). This latter
quantity is calculated from the heliocentric distance of the asteroid,
the value assumed for the albedo, and the solar incident
angle. Macroscopic surface roughness is usually modelled by adding
hemispherical section craters of variable opening angle and variable
surface density to each facet. Shadowing and multiple reflections of
incident solar and thermally emitted radiation inside craters are
taken into account as described by
\cite{Spencer1990Icar...83...27S,Emery1998Icar..136..104E,Rozitis2011MNRAS.415.2042R}
and \cite{Lagerros1998A&A...332.1123L}. Heat
conduction is also accounted for within craters
\citep{Spencer1989Icar...78..337S, Spencer1990Icar...83...27S,
  Lag:96a, Delbo2004PhD}. Surface roughness can be
adjusted by changing the opening angle of the craters, the density of
the crater distribution, or a combination of both
\citep{Mueller2007PhD}. 
The total observable thermal emission is calculated by summing the
contributions from each facet visible to the observer. Model
parameters are adjusted until the best agreement with observational
data  is obtained, i.e., the least-squares residual of the fit is
minimized, thereby constraining the physical properties (albedo, size,
macroscopic roughness, and thermal inertia) of the asteroid. 

From the point of view of multi-data inversion however, the optimization of thermophysical parameters as described above is a two-step process -- first, the spin and shape model is derived from one data type (photometry, radar,\ldots), then this model is fixed and used for deriving thermophysical parameters from another data type (thermal infrared). This approach lacks the possibility to weight the two data types with respect to each other. Moreover, the thermophysical parameters can be very sensitive to small modifications of the input shape and spin, so various modifications of the shape should be tested to see how stable the solution is \citep{Han.ea:15}. Ideally, one should  
model shape and spin parameters together with
thermal parameters. This multi-data approach using lightcurves and
thermal infrared data simultaneously was successfully tested by \cite{Dur.ea:12, Dur.ea:14} and in principle it can be used also to data that are sparse in time. 

 \subsection{\textbf{Radar}}
 
\begin{figure}[t]
\centering
\reflectbox{\includegraphics[clip=true,trim=70 100 70 90,scale=0.55]{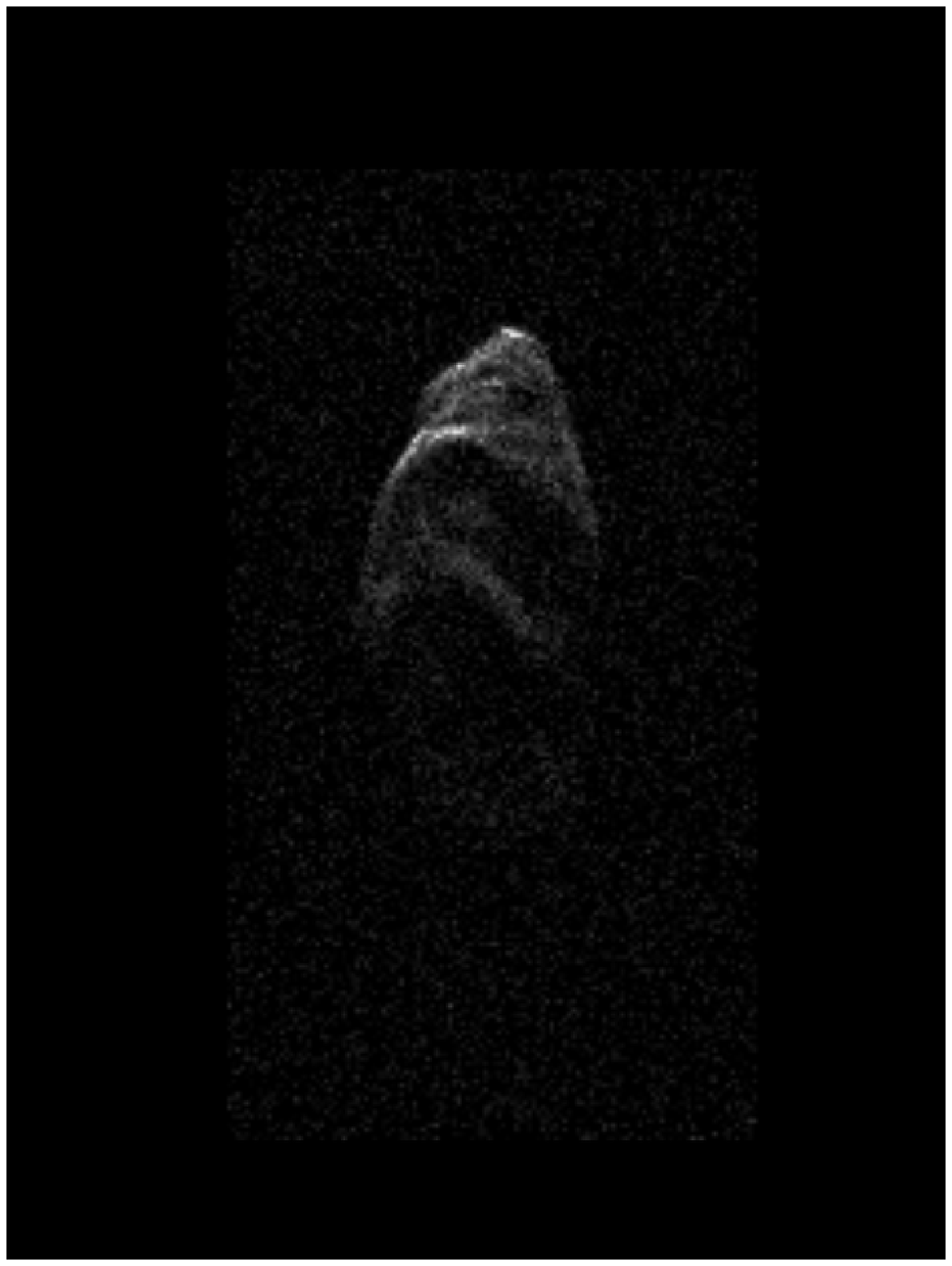}}
\reflectbox{\includegraphics[clip=true,trim=70 110 70 80,scale=0.55]{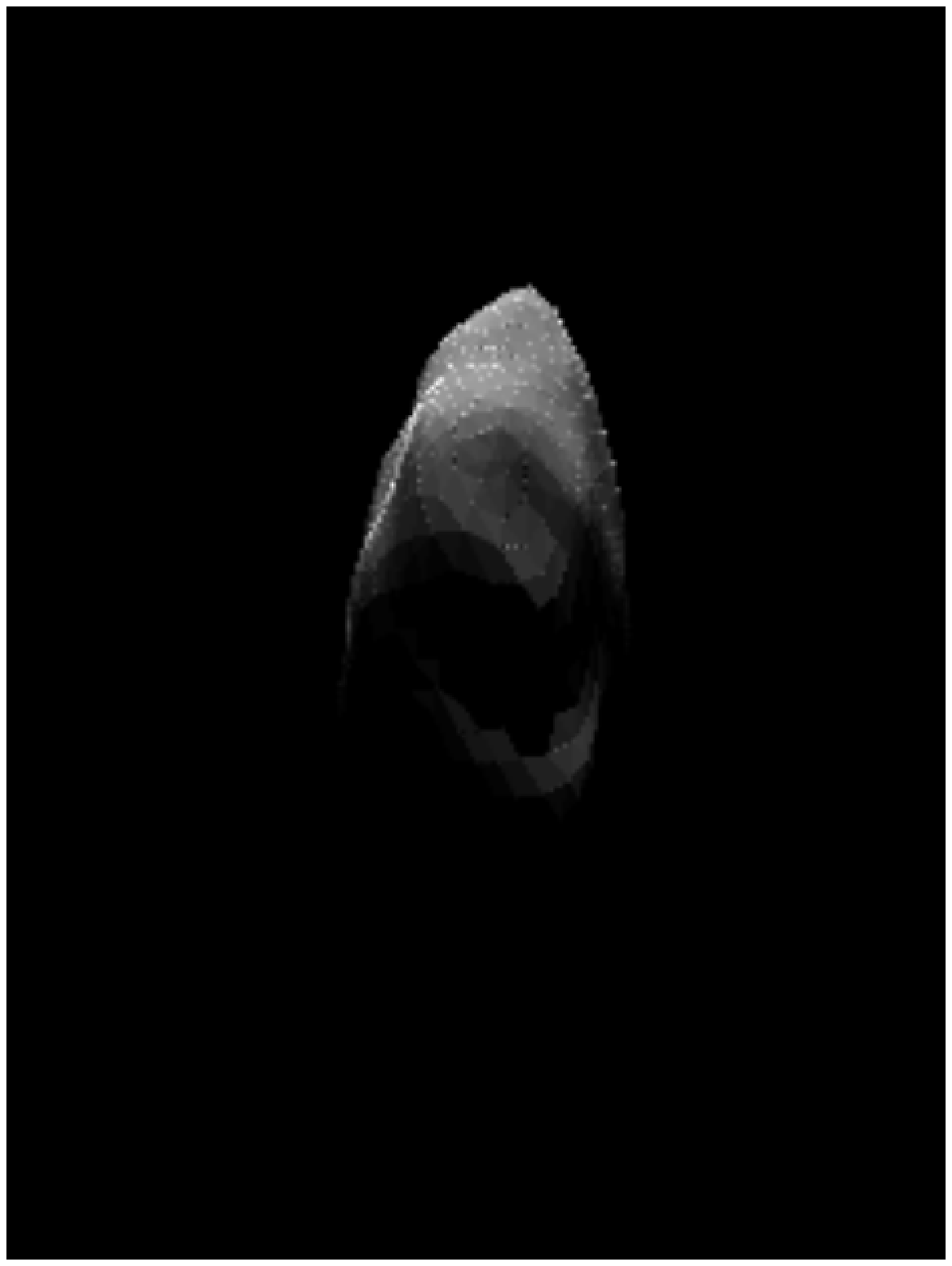}}
\reflectbox{\includegraphics[clip=true,trim=70 100 70 90,scale=0.55]{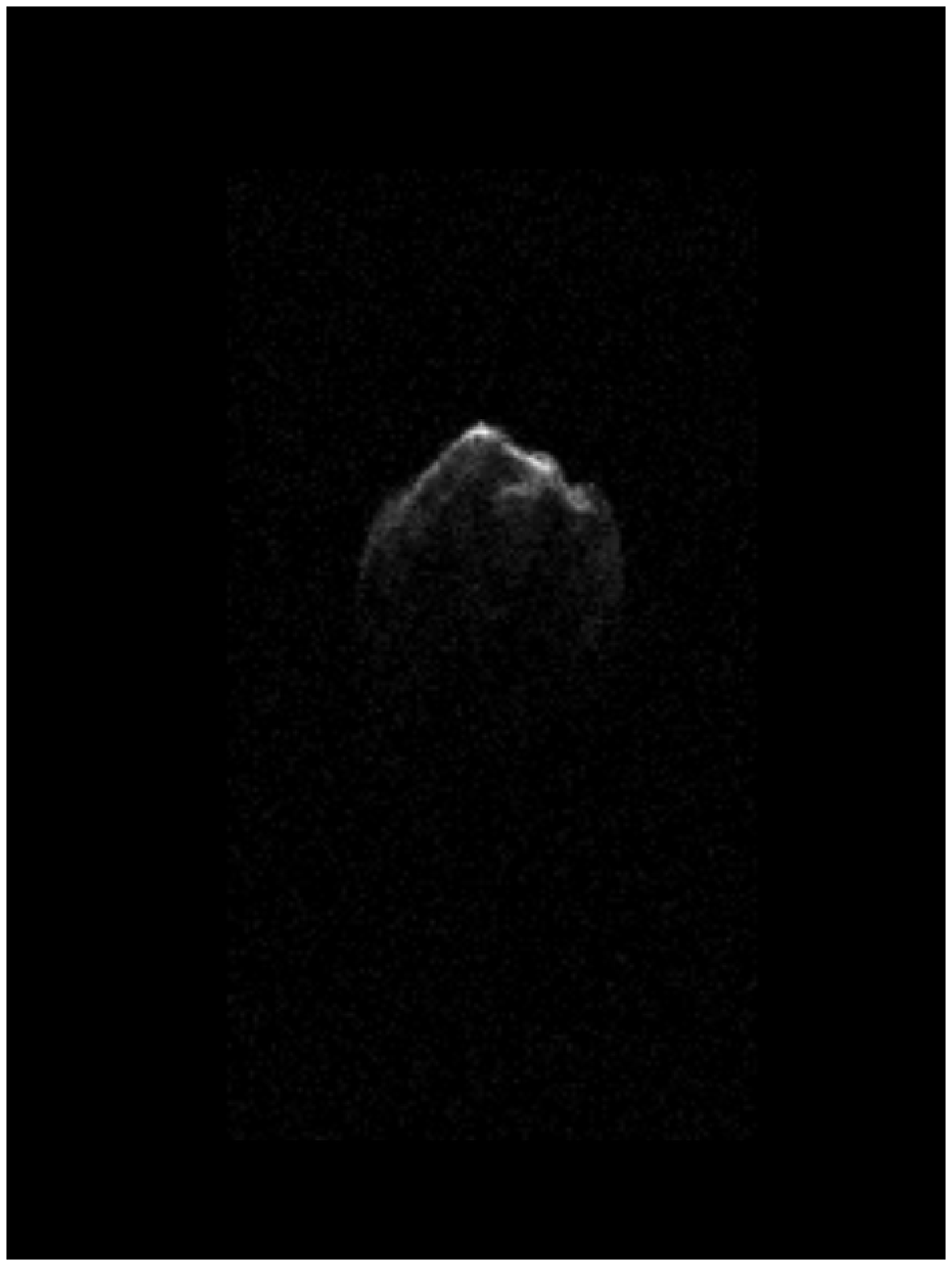}}
\reflectbox{\includegraphics[clip=true,trim=70 110 70 80,scale=0.55]{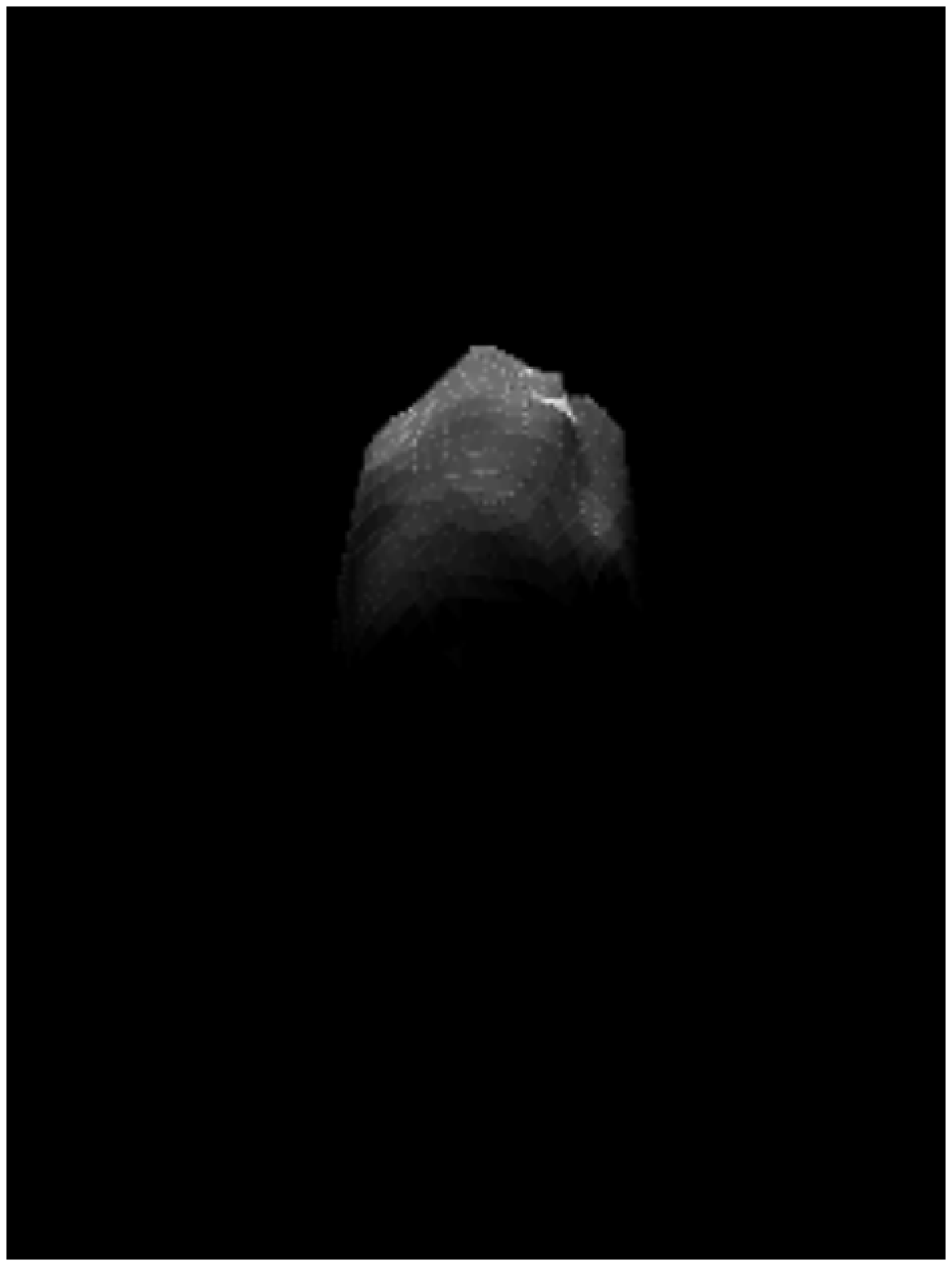}}
\caption{\small Example range-Doppler radar images (left) of asteroid
2000~ET$_{70}$ observed at Arecibo observatory \citep{Nai.ea:13}. Range
and frequency resolutions are $15$\,m and $0.075$\,Hz, respectively. Range increases towards the bottom, Doppler frequency increases to the right. 
The simulated images (right) correspond to the reconstructed shape model
in Fig.~\ref{fig:radar_shape}.}
\label{fig:radar_data}
\end{figure}

\begin{figure}[t]
\epsscale{1}
\plotone{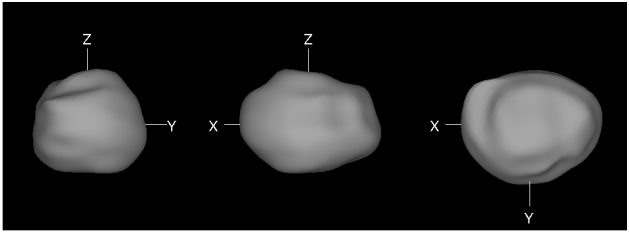}
\caption{\small Shape model of the asteroid 2000~ET$_{70}$ reconstructed from Arecibo and Goldstone delay-Doppler radar images with ADAM (Sect.~\ref{sec:software}).}
\label{fig:radar_shape}
\end{figure}

  Radar observations that measure the distribution of echo power in
  time delay and Doppler frequency (so-called range-Doppler or 
  delay-Doppler measurements) are discussed in detail in \cite{Ost.ea:02} and in
  {\it Benner et al.} (this volume). The delay-Doppler projection is many-to-one mapping of a 3-D surface of the target into a 2-D ``image''. Each pixel on the image represents a bin containing integrated echo power from surface elements that have the same distance from the radar and the same relative speed (due to the rotation of the asteroid). From the point of view of 
  inversion, images in the range-Doppler plane are generalized
  projections that can automatically be handled with the general
  procedure discussed in Sect.~\ref{sec:inversion} and in detail in
  \cite{Vii.Kaa:14} and \cite{Vii.ea:15}. In this approach, the multi-mode
  reconstruction is tuned to produce models with intermediate scale
  resolution ($\sim$1/10 of the diameter) since these are computationally inexpensive (can be obtained
  in a few minutes with a laptop), and data sources other than radar do
  not contain more detailed information. An example of radar
  range-Doppler data of asteroid 2000~ET$_{70}$ is shown in
  Fig.~\ref{fig:radar_data} and the corresponding reconstructed shape
  model in Fig.~\ref{fig:radar_shape}. If detailed radar data are 
  available, such model can then be further refined \citep{Nai.ea:13}
  with the radar 
  techniques described by {\it Benner et al.} (this volume).

Due to the steep decrease of echo power with the distance to the
object (the fourth power of the distance), asteroids achievable by
current radar facilities Arecibo and Goldstone are only
close-approaching near-Earth asteroids or the largest members of the
main belt. A number of examples of what can be reconstructed from radar data is given by {\it Benner et al.} (this volume).

 \subsection{\textbf{Asteroid interior}}
 
    We briefly discuss here the interesting possibility of peer into an asteroid. This is somewhat separate from
    the remote-sensing framework as the data can only be obtained in
    situ; on the other hand, such data can be acquired with the future
    space missions. The most practical option is to plant radio
    transmitters/receivers on the surface of a kilometer-sized
    asteroid and measure low-frequency ($\sim$\,100\,MHz) signals
    between these and an orbiter when they pass through the interior
    of the target.  

The most robust observables are simply signal travel-time data \citep{Pur.Kaa:13}. These allow an efficient formulation of the inverse problem via the refraction index and are relatively insensitive to noise and model error. Nevertheless, they suffice to give a coarse-scale picture of the general distribution of permittivity inside the asteroid, as well as the locations and sizes of large anomalies (sudden low- or high-density regions such as voids or heavier minerals). This approach has also been robustly tested in laboratory  conditions \citep{Pur.Kaa:14b}. A more refined possibility is to measure changes in the pulse profile, although this is more prone to errors \citep{Pur.Kaa:14a}. The interior of the asteroid is practically impossible to model accurately in three dimension since it is supposed to have a number of cracks, voids, discontinuities etc., all refracting and reflecting the radio waves in complicated ways. Thus a very robust scheme is essential for extracting the available information with stability. 
Regardless of the data type, one or two transmitters on the surface are not sufficient for a unique solution. A tetrahedral configuration of four transmitters would be ideal, but this places heavy demands on the payload design.

\subsection{\textbf{Extension of the model}}

So far, we assumed that the asteroid can be described as a solid single body with constant spin vector, i.e., rotating along the axis with the maximum moment of inertia with a constant rotation rate. Although this model represents a typical asteroid, there are other configurations that can be also treated with an extension of the simple model. 
 
\subsubsection{Binaries}

\begin{figure}[t]
\epsscale{0.8}
 \plotone{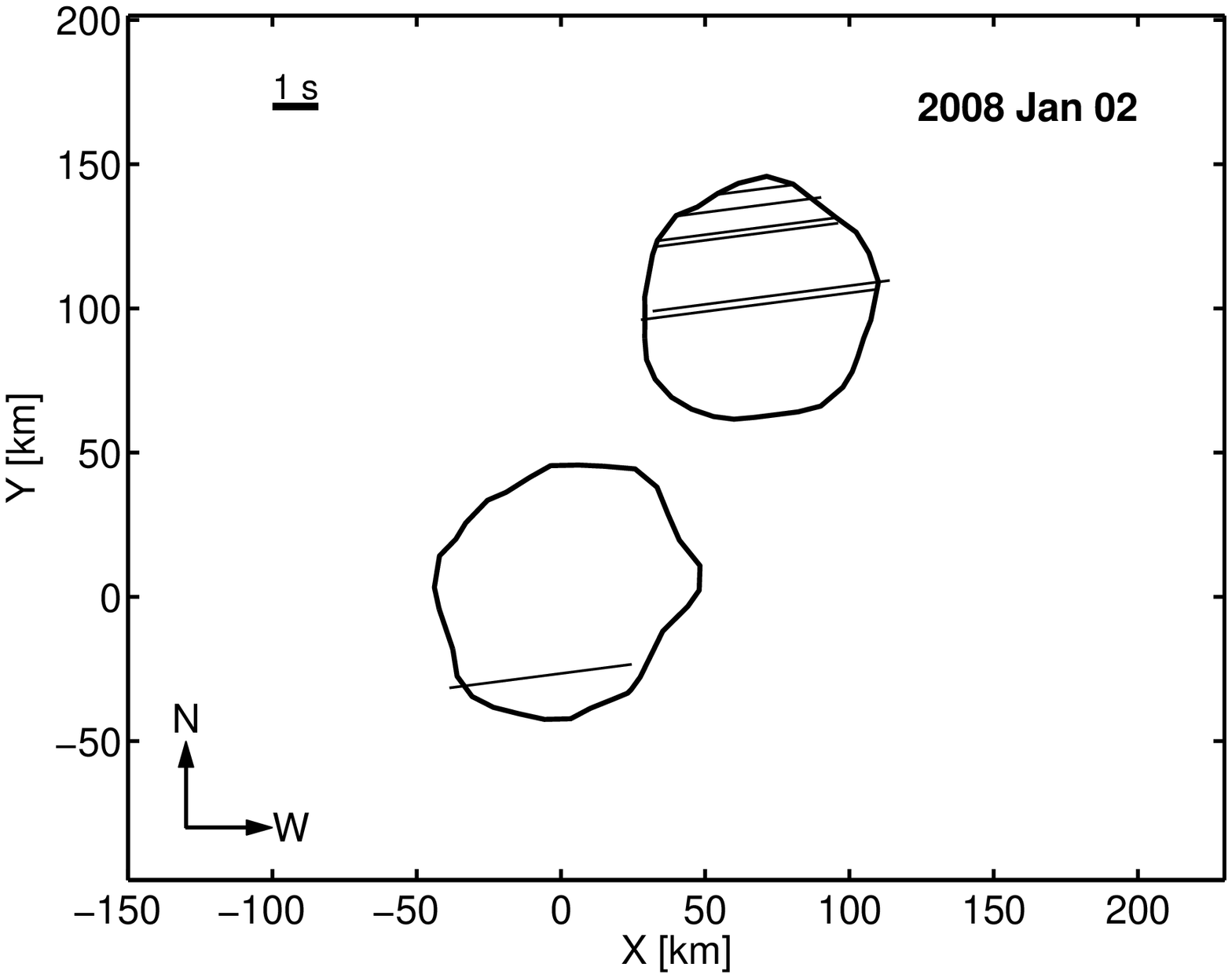}
 \plotone{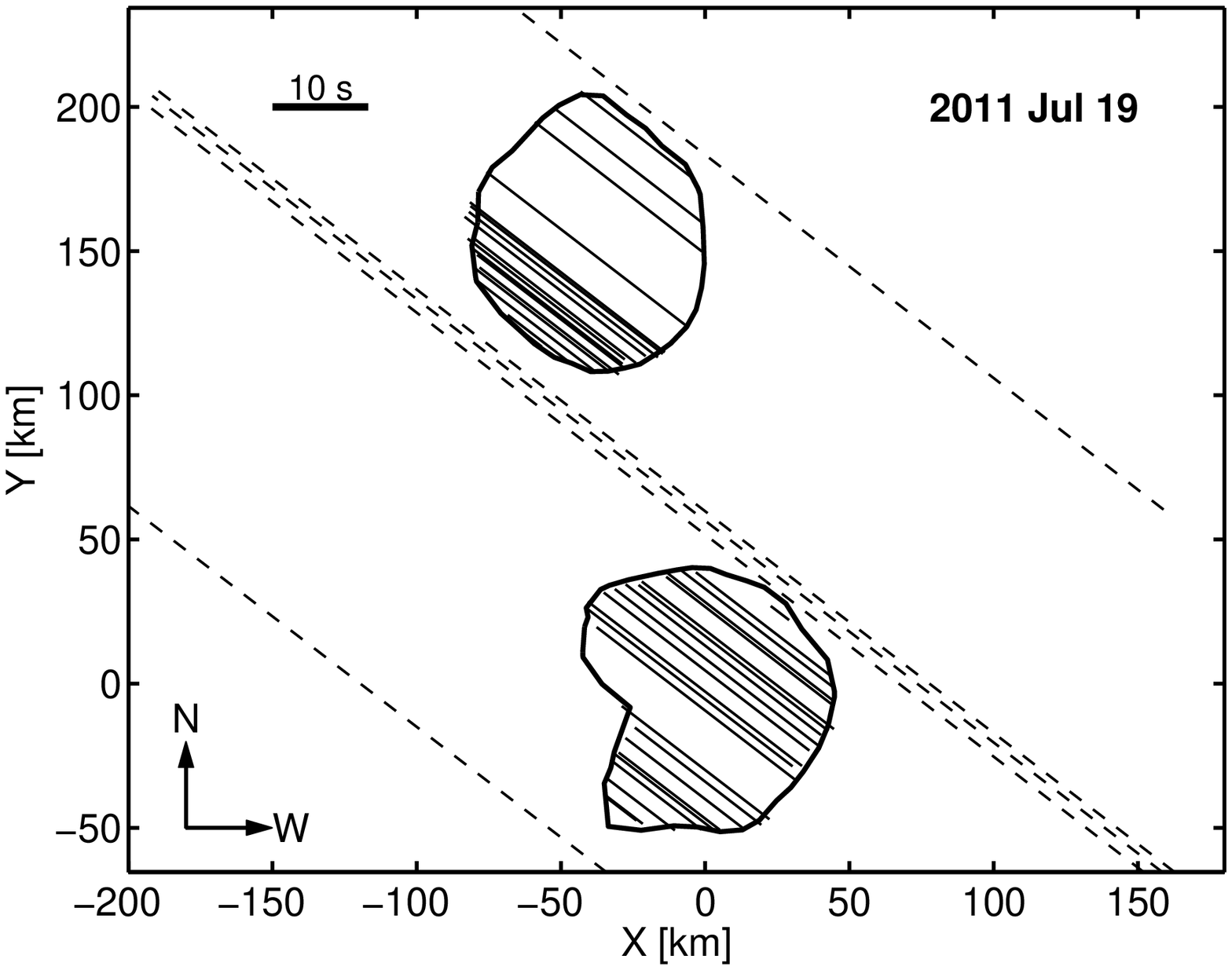}
 \plotone{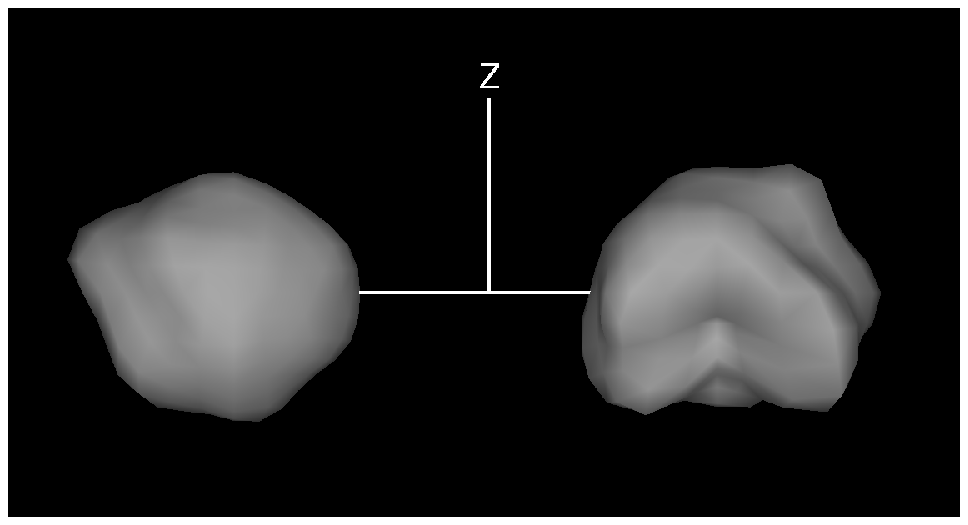}
 \caption{\small Shape model  of (90)~Antiope seen form its equator (bottom) reconstructed from lightcurves and two occultations observed in 2008 and 2011 (top and middle). The model silhouette is projected on the plane of the sky, the solid lines are positive chords, the dashed ones are negative observations.}  
 \label{fig_Antiope}
\end{figure}

  Binary asteroids form a significant part of the population of small 
  asteroids. \cite{Pra.ea:06} estimated that the fraction of binaries is 15\% for the near-Earth population and a similar fraction is assumed for the main-belt population in the same size range ({\it Margot et al.}, this volume).
  The formation, dynamics, and physical properties of binary and multiple asteroid systems are discussed in detail in other chapters of this volume ({\it Walsh et al., Margot et al.}). 
  In general, modeling of such systems is more complex because the dynamics has to be taken into account. However, in some cases described below, the technique used for modeling 
  single asteroids can be used also for multiple systems or at least their primary components. 
  
  For binary (or multiple) systems where the primary component is much larger than
  the satellite, the photometric signal from the satellite can be neglected and
  the primary can be modeled as a single body. For such systems, the
  shape of the primary and the constraints on its 
  gravitational quadrupole
  $J_2$ from the orbit analysis of the secondary can be used to investigate the distribution of the
  density \citep{Ber.ea:14, Tak.Sch:14}. 
  
  For systems with comparable sizes, the problem becomes complicated when
  the system is asynchronous, i.e., when the rotation period of the
  primary is different from the orbital period of the secondary. Such
  systems are  usually modeled as two ellipsoids
  \citep{Sch.Pra:09}, although more general models were created from radar
  observation \citep[1999 KW$_4$, for example, see][]{Ost.ea:06}. 

Fully synchronous binaries can be approximated by single bodies if the separation of their components is not large. Even a convex model can provide a good fit to the lightcurves \citep{Dur.Kaa:03}. Such model does not represent the true configuration of the system, of course, but it provides the correct rotation period and direction of the orbital plane. 

When the separation of components is larger, the system has to be modeled as
two-component. However, from the modeling point of view, it is just a
moderate modification of the nonconvex problem, where the system is
described by only one rotation/orbital period and orientation of the
normal of the orbital plane (parallel to the spins of the bodies). If
the model is based on lightcurves only, the spin and period parameters
can be reconstructed accurately, but the uncertainty in shapes is
large. As has been shown by \cite{Mar.ea:14} on (624)~Hektor, the
distinction between a highly nonconvex single body, two bodies in
contact, or two bodies orbiting each other is difficult to make.  

An example of reconstruction of a doubly synchronous binary
system (90)~Antiope from lightcurves 
and occultations is shown in Fig.~\ref{fig_Antiope},
where the model is shown together with the silhouettes from
occultations. Tens of chords observed during the occultation in 2011 \citep{Col.ea:12} portray
the two components to details unattainable by any other observational
technique and the large set of lightcurves observed over many apparitions constrains the rotational
parameters. A similar model can be obtained also by using lightcurves separately to create a scale-free model that is then scaled by occultation data \citep[the Shaping Asteroids with Genetic Evolution algorithm,][]{Bar.ea:14}. 
However, this two-step approach lacks the advantages of simultaneous
inversion where the two data types can be weighted
with respect to each other. 
 
\subsubsection{YORP effect}

As described in detail by {\it Vokrouhlick\'y et al.} (this volume),
rotation state of small asteroids is affected by the anisotropic
recoil of scattered sunlight and thermal radiation, which causes a net torque called the
Yarkovsky-O'Keefe-Radzievskii-Paddack (YORP) effect. This effect
secularly changes the obliquity of the spin vector and the rotation
period. Whereas the former is too small to be measured with current
data, the latter has been measured on several asteroids (see Table~3 in
{\it Vokrouhlick\'y et al.}, this volume). If the change of the
rotation period is larger than the uncertainty of the period, the
change can be traced from apparition to apparition as was the case for
(54509)~YORP \citep{Low.ea:07}. In other cases, the effect was much
smaller and it revealed itself by the discrepancy between the data and
the model assuming the period to be constant \citep{Kaa.ea:07,
  Dur.ea:08b, Dur.ea:12b, Low.ea:14}.  

The YORP effect is easy to include into the model. We assume that the rotation rate $\omega$ changes linearly in time $t$ as $\mathrm{d}\omega / \mathrm{d} t  = \upsilon$. Then the parameter $\upsilon$ is another free parameter of the modeling. Because the shift in the rotation phase increases quadratically in time as $\phi = \omega t + 1/2 \upsilon t^2$, even  small changes $\mathrm{d}\omega / \mathrm{d}t$ of the order $10^{-8}$\,rad/d$^{-2}$ can be detected with data sets covering tens of years. 

In principle, the measured value of $\upsilon$ can be compared with the theoretical value computed from the spin state, shape, size, and the thermal parameters of the surface with the density as a free parameter. However, due to sensitivity of YORP on small scale details of the shape that are far below the resolution of the model \citep{Sta:09, Kaa.Nor:13}, and the problem of transverse heat diffusion \citep{Gol.Kru:12, Gol.ea:14, Sev.ea:14}, this can hardly be more than a rough comparison.

Because the YORP effect scales as inverse of the square of the size of asteroid, it becomes more important for small bodies, where it might be necessary to include it into the modeling if the data cover a wider span of time. YORP is assumed to play an important role in many dynamical processes -- the distribution of rotation periods and spin obliquities of small asteroids \citep{Pra.ea:08, Han.ea:13b} or the creation of asteroid binaries and asteroid pairs \citep{Pra.ea:10}, for example. It is important to have more asteroids with YORP detection. Better statistics of values of period change will help to constrain theories of YORP evolution of small asteroids.
 
\subsubsection{Excited rotation}

Asteroids rotating in the relaxed mode are fully described by the spin
axis direction, rotation rate, and the initial orientation. However,
some asteroids are in an excited rotation state, which can be described
as a rotating free top. More parameters are needed to describe this
tumbling motion \citep{Kaa:01}. The reason why some asteroids are in
this state can be (i) primordial, (ii) collisional excitation
\citep{Hen.Pra:13}, or (iii) end state of YORP-driven spin down. The
approach to the modeling is in principle the same as for asteroids in
principal axis rotation, only the orientation for a given time is
given by solving differential equations. From lightcurves, models of
asteroids 2008~TC$_3$ \citep{Sch.ea:10} and (99942)~Apophis
\citep{Pra.ea:14} have been derived. A model of asteroid (4179)~Toutatis
\citep{Hud.ea:03} was derived from radar data and lightcurves. This
model was later compared with the fly-by images of Chang'E-2 mission
-- the general shape was in agreement, although there were some minor
discrepancies \citep{Zou.ea:14}. 

\subsection{Procedures: Convexinv, KOALA, and ADAM}
\label{sec:software}

Software for the inverse problems is available at
DAMIT (\url{http://astro.troja.mff.cuni.cz/projects/asteroids3D}).
 {\tt Convexinv} is a procedure for lightcurve inversion, while ADAM
(All-Data Asteroid Modeling) is a collection of functions from which
one can tailor an inversion procedure for any data sources
\citep{Vii.ea:15}. An earlier version of this is called KOALA
\citep[Knitted Occultation, Adaptive optics and Lightcurve
  Analysis,][]{Car.ea:12}; this is based on lightcurves and silhouette
contours obtainable from images and occultations
\citep{Kaa:11,Kaa.Vii:12}. KOALA is especially suitable for lightcurve
and occultation data, while ADAM allows the use of any images
(camera, radar, or interferometry) with or without lightcurves without
having to process them to extract contours or other information. 
Both KOALA and ADAM can be used for lightcurves only but, as discussed earlier, this is not reliable. Any resulting shape
should only be taken in the global sense (as a more realistic-looking rendering of a convex solution), and the details or
nonconvex features are seldom likely to be real \citep{Vii.ea:15}.

\section{\textbf{WHAT HAVE WE LEARNED FROM 3-D SHAPES?}} \label{sec:results}

\begin{figure}[t]
 \epsscale{1}
 \plotone{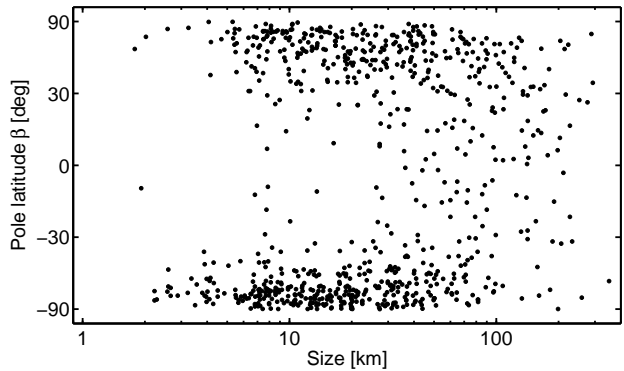}
 \caption{\small Distribution of pole ecliptic latitude $\beta$ with
   respect to the size of the asteroid for $\sim\,800$ asteroids. The the scale on the vertical axis is linear in $\sin\beta$, which makes the vertical distribution of points in this plot uniform for an isotropic distribution of spin vectors.}   
 \label{fig_beta_vs_size}
\end{figure}

The purpose of modeling methods described in previous sections is to reveal new facts about the nature of asteroids. The approach is made on two fronts: The first strategy is to use the most abundant data sources (photometry in visual and thermal IR) to produce many low-resolution models that will be a statistically significant sample of the whole asteroid population from which conclusions about the physical properties can be drawn. The second approach is to concentrate on selected targets, obtain as many different data types as possible, create detailed models of these asteroids, and extrapolate the obtained results to the whole class of similar objects.
  We describe in this section new research areas that directly benefit
  from availability of spin solutions and 3-D shape models.

\subsection{\textbf{Spin-axis distribution and evolution}}
\label{spin_distribution}

One of the main results of the lightcurve inversion is the increasing
sample of asteroids with known orientation of the spin axis. For main
belt asteroids, the long-known lack of asteroids with poles close to
the ecliptic \citep{Kry.ea:07} was confirmed and it was shown that it
is more pronounced for smaller asteroids \citep{Han.ea:11}. In
Fig.~\ref{fig_beta_vs_size}, we plot the distribution of pole
latitudes for $\sim$\,800 asteroid models with respect to their size
\citep[an updated version of Fig.~5 in][]{Han.ea:11}.
The size-dependent structure can be explained by the YORP effect that
is more effective on smaller asteroids ($\lesssim$\,30\,km) and pushes
them into extreme values of obliquity (0$^\circ$ or 180$^\circ$). This
corresponds to the clustering of pole latitudes towards values of
$\pm$\,90$^\circ$. Although there are observation and modeling biases that affect the distribution of poles in the sample of available models, their effect is only marginal compared to the strong anisotropy seen in Fig.~\ref{fig_beta_vs_size} \citep{Han.ea:11}. The spin-axis orientation of even smaller asteroids
($\lesssim$\,5\,km) is still not known due to the lack of models. For
the largest asteroids ($\gtrsim$\,60\,km), there is a statistically
significant increase of prograde rotators (98 prograde vs. 63 retrograde in Fig.~\ref{fig_beta_vs_size}), probably of primordial
origin \citep{Joh.Lac:10}. 
 
A different approach to the problem of spin-axis distribution was used by \cite{Bow.ea:14}. They analyzed variations of the mean brightness with the ecliptic longitude, from which they estimated ecliptic longitudes of spin axis for about 350,000 asteroids and revealed a clearly non-uniform distribution. However, the explanation of the cause for this non-uniformity is still missing.

\begin{figure}[t]
 \epsscale{1}
 \plotone{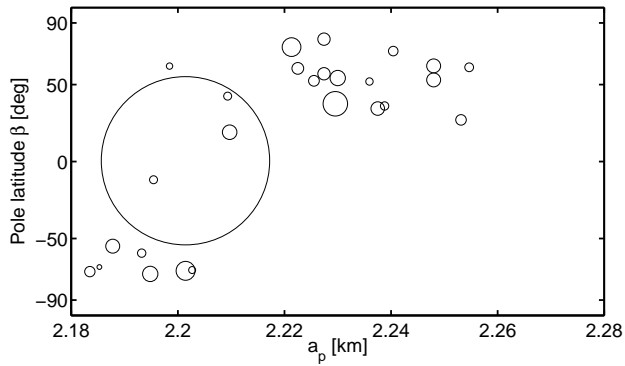}
 \caption{\small Distribution of pole ecliptic latitude $\beta$ with respect to the proper semimajor axis $a_\mathrm{p}$ of the Flora family members. The relative size of each circle corresponds to the size of the asteroid.}  
 \label{fig_Flora}
\end{figure}

With increasing sample of models, it is also possible to study the distribution of spin axes of members of collisional families. Results of \cite{Han.ea:13a} agree with theoretical expectations: if the spread in proper semimajor axis increases with decreasing size due to the Yarkovsky effect ({\it Nesvorn\'y at al.}, {\it Vokrouhlick\'y et al.}, this volume), asteroids closer to Sun than the center of the family should rotate retrograde, those farther should rotate prograde. This behavior is shown in Fig.~\ref{fig_Flora} for Flora family.

  \subsection{A pre-requisite in many cases}
  Unlike most astronomical objects, the viewing geometry of asteroids
  continuously changes due to their motion relative to observer.
  As a corollary, knowing their rotation period and spin-vector
  coordinates is crucial to
  interpret projected size measurements, and
  to tie together observations.

  As already described in Sect.~\ref{sec:data}, the spin and 3-D shape
  are required to interpret 
  thermal infrared radiometry, stellar occultations, or apparent sizes measured by Gaia without biases.
  Similarly, there have been long running discussions to explain
  inconsistent spectral measurements of asteroids, that can be easily
  solved once the spin properties (period and orientation) are known:
  see, for instance, the discussion on (832)~Karin by 
  \citet{2004-ApJ-615-Sasaki}, 
  \citet{2007-Icarus-191-Vernazza}, and
  \citet{2007-Icarus-191-Chapman}, or 
  on (21)~Lutetia by \citet{2012-PSS-66-Barucci}.\\

  \subsection{Density, composition, and internal structure}
  Density is one of the most fundamental properties to constrain the
  composition of asteroids and investigate their internal structure
  ({\it Scheeres et al.}, this volume).
  With the exception of binary asteroids with observable 
  mutual eclipses ({\it Margot et al.}, this volume), both mass and
  volume are required to determine the density of an asteroid.
  Estimating the mass of an asteroid by measuring its gravitational influence on other objects is a challenge because of the relative
  low mass of asteroids compared to other planetary objects.
  The number of mass determinations thus limits the number of density
  estimates.
  Although there are diameter, hence volume, estimates for all
  asteroids with a mass determination, the uncertainty in volume generally
  dominates the balance on density uncertainty
  \citep{2012-PSS-73-Carry}. 
  
  In this respect, 3-D shape models are required to determine accurate
  volume. The level of potential biases increases with stronger
  assumptions on the shape, and accuracy accounting for systematics
  improves from sphere, to ellipsoid, to convex hull, to the real
  shape. 
  In the decade since {\em Asteroids III\/}, the number of density estimates
  has increased from 20 to 300 objects
  \citep{2002-AsteroidsIII-4.2-Britt, 2012-PSS-73-Carry}.
  Among these, the most reliable are derived from binary systems for
  which the volume of the primary was determined after 
  shape modeling
  \citep[e.g.,][]{Ost.ea:06, 
    2006-Icarus-184-Shepard, 
    2009-Icarus-203-Descamps, 2011-Icarus-211-Descamps,
    2011-AJ-141-Fang, Mar.ea:13}.

  A detailed description on the density of asteroids and their
  internal structure can be found in {\it Scheeres et al.} (this volume),
  from under-dense asteroids, hosting large voids, to over-dense
  asteroids, likely differentiated.
  In the context of this chapter, it is important to highlight than only density estimates more
  precise than 10--20\% can be used to discriminate between different
  analogue meteorites and can provide insights on the internal
  structure.
  Such accuracy can only be achieved with volume known to 5--10\%
  or better, 
  which means that a proper description of the 3-D shape is needed.

\subsection{Determination of surface properties by means of TPMs}
The spin state and shape model of an asteroid is input information for the TPMs. While in the {\it Asteroids III}
era, shape and spin information were available only for a handful of asteroids, preventing application of TPMs to a large
number of these bodies, this situation has drastically changed in the last few years. Physical properties (such as the value
of the surface thermal inertia) of about 60 asteroids are now available thanks to the application of TPMs (see {\it Delbo et
al.}, this volume). A remarkable improvement in this field is also represented by the availability of high quality thermal
infrared data as those produced but the WISE and the Spitzer space telescopes. In the next few years we expect the number
of asteroids with known thermal inertia values to grow thanks to the availability of more shape and spin state models. 

  \subsection{Surface re-arrangement}
  From the spin, 3-D shape,
  and density, the local gravity at the surface can be computed.
  Unsuspected physics has been unveiled with the modeling of the
  near-Earth asteroid (66391) 1999 KW$_4$ by
  \citet{Ost.ea:06}.
  Some small asteroids present an equatorial bulge, presumably generated
  by regolith migration toward lower gravity regions. This process
  can even form binary systems if the asteroids spin fast enough
  \citep{2008-Nature-454-Walsh,2009-Icarus-199-Harris}. \\

  \subsection{Cratering events} 
  An evident outcome of shape
  modeling is the capability to detect large impact craters and
  basins. 
  Aside from the spacecraft encounters, the first detection was the
  large impact basin on Vesta, progenitor of the Vestoids
  \citep{1993-Science-260-Binzel}, detected with the HST
  \citep{1997-Science-277-Thomas} and confirmed by NASA Dawn spacecraft
  \citep{2012-Science-336-Russell}.
  Another case is the recent impact suffered by asteroids (596)
  Scheila, detected by the presence of a dust tail
  \citep{2011-ApJ-733-Bodewits}. Lightcurves obtained before and after
  the impact, under similar geometries are different, revealing
  different surface properties \citep{2014-Icarus-229-Bodewits}. \\

\begin{deluxetable}{lcccccrrr}
\tabletypesize{\small}
\tablecaption{\label{tab:tech} A list of observation techniques and derivable physical properties.}
\tablewidth{0pt}
\tablehead{Technique & Period & Spin & Size & Shape & Thermal Inertia & \multicolumn{3}{c}{Number of models}\\ 
\cline{7-9}
	             &        &      &      &       &                 & \multicolumn{1}{c}{Asteroids III} & \multicolumn{1}{c}{Asteroids IV} & \multicolumn{1}{c}{Asteroids V}\\ }
\startdata
Photometry  	& X & X &   & X &    &	30		& 500		& $10^4$   	\\
Images      	&   & X & X & X &    &	5		& 50		& $10^2$	\\
Occultation 	&   & X & X & X &    &	5$^\ast$	& 50		& $10^2$	\\
Radar       	& X & X & X & X &    &	10		& 30		& $10^2$	\\
Radiometry  	&   &   & X &   &  X &	10$^\ast$	& 20		& $10^4$	\\
Interferometry  &   & X & X & X &    &	5$^\dag$	& $<10$		& $10^2$	\\
Fly-by		& X & X & X & X &  X &	6		& 10		& $<15$		\\
\enddata
\tablenotetext{\ }{The ``X'' mark indicates which physical properties are derivable from which technique. The number of models available at the time of {\it Asteroids III} book (Asteroids III) and now (Asteroids IV) is only approximate. The Asteroids V column is an order-of-magnitude estimate for the next decade.}
\tablenotetext{\ast}{Ellipsoidal models.}
\tablenotetext{\dag}{HST/FGS.}
\end{deluxetable}
    
  \subsection{Mass distribution}
  Because some asteroids are
  less or more dense than their most-likely constituents, the question
  of the mass distribution (denser material or voids) in their
  interior can be asked. This question is intrinsically tied with the
  study of the gravity field around the asteroid.
  The latter has been measured during spacecraft encounters
  \citep[see][for instance]{2002-Icarus-155-Miller}, but studies from
  Earth-bound observations have recently appeared.
  By comparing the spherical harmonics of the gravity field as
  determined from the orbit of a natural satellite, with the expected
  coefficients resulting from the 3-D shape model, the hypothesis of
  homogeneous mass distribution can be tested
  \citep{2012-AA-543-Vachier, Ber.ea:14,
    Tak.Sch:14}.

\section{\textbf{FUTURE}}
\label{sec:future}

In the decade since {\it Asteroids III}, where the principles of
    lightcurve inversion based on dense-in-time series were presented
    \citep{Kaa.ea:02}, the number of models has seen a tenfold increase -- from a few
    tens to a few hundreds (Table~\ref{tab:tech}). 
    The increasing availability of sparse-in-time photometry, and its
    appropriate handling in the inversion, coupled with a dramatic increase of computer time (thanks to projects like Asteroids@Home)
    have made this possible.
    Upcoming all-sky surveys such as Pan-STARRS, LSST, and Gaia are
    expected to produce enormous data sets and there is little doubt
    that thousands of models will be derived in the next decade.
    Our knowledge on non-gravitational effects such as YORP and
    Yarkovsky will directly benefit from this larger sample.

With the large number of data and models, new challenges will arise -- how to extract scientifically interesting information from a large set of models of asteroids? With big data flows, the processing has to be automated, with effective data processing. The obvious search for correlations has to be done with care because of large biases in the set of models. Although the importance of detailed models of individual well-studied asteroids will be important, the main shift in paradigm and probably the main source of interesting findings will be in tens of thousands of asteroid models derived from photometry in optical and thermal infrared wavelengths.   Only a few years ago, only photometry in the visible was available
    for a large number of asteroids. The situation drastically changed
    with WISE catalog of thermal fluxes for 150,000 asteroids.
    Automatized procedures capable of dealing with photometry in the
    visible and thermal infrared will yield not only 3-D shape and
    spin state for thousands of asteroids, but also their diameter,
    albedo, and thermal inertia of their surface. The later being
    crucial in interpreting asteroid mineralogy once coupled with
    spectroscopy ({\it Reddy et al.}, this volume). 
Understanding of observational and modeling biases will be crucial for correct interpretation of the results. Connecting spin and shape distribution of asteroids with their orbital and spin evolution will hopefully lead to a clear picture of the evolution of the main asteroid belt. The ultimate goal here is the connection of models of evolution of the 
main belt with spin-axis evolution and current distribution of asteroid physical properties.

Another approach that is complementary to modeling individual objects,
is modeling distribution functions of parameters of interest. With any inversion technique and photometric data quality, the number of models will be always much lower than the number of known asteroids just because it takes time to collect enough data at different geometries. So instead
of aiming to create unique models for a statistically significant sample of the (sub)populations with known observational and modeling bias, one can use other observables than the
time--brightness pairs. With this approach, there is almost no
``wasting'' of data because essentially all photometric points are
used in the statistical approach. The aim is to model characteristics
of a given asteroid population as a whole when there
are not enough data to model individual members. 
One of the observables can be, for example, the mean brightness over
one apparition \citep{Bow.ea:14} and its dispersion that
correlates with lightcurve amplitude. If there are not enough data
points to estimate mean brightness and its variance, the statistics of
scatter of individual pairs of observations can be used \citep{Sza.Kis:08}.  
If the data
come in pairs of observations separated by a constant time interval
(like with Pan-STARRS), the rotation period can be estimated
\citep{Dur.ea:07}. 

  Another challenge for next decade resides in the derivation of a
    large sample of higher-resolution models, based on multiple
    data sources. Building inversion techniques capable of handling
    different data sources had been highlighted in {\it Asteroids III} as
    the next step, and this became reality over the past few years.
    The models based on multiple data sources are more realistic,
    more reliable, more precise than the independent analysis of
    individual data sets. If practical issues of merging data sets
    arise, mathematical solution have been exposed, and algorithms and
    software are ready and freely available. 
    The current sample of models derived by such methods is
    nevertheless still limited. 
If the multi data analysis is more efficient than the independent analysis of the individual data sets, the practical problem of data sharing arises. Although the tools to invert multiple data sources are ready and freely available, the data are still hard to get and harder to share. To maximize the scientific output, it is necessary to combine all available data sources.

    The other observing techniques (i.e., disk-resolved imaging, 
    stellar occultations, Gaia size measurements, or radar echoes)
    will never provide data for
    more than a few thousands asteroids.
    The apparent sizes by Gaia will however be measured with a high
    precision, and will concern a decent sized sample. Because the
    data will be made public, they should be used on a regular basis
    for modeling.
    The number and quality of profiles derived from stellar occultations have
    always been restricted due to the intrinsic complexity of
    prediction, and cost of equipment. The parallel availability of
    low-cost telescopes and cameras, together with the publication of
    Gaia stellar catalog and asteroid orbits, will open a new age for
    stellar occultations.
    Any mass-production procedure of asteroid models should therefore
    be able to deal not only with photometry (visible and infrared)
    but also with stellar occultations.  

  Finally, if the bulk of asteroids are single objects,
    rotating along their shortest axis, a small fraction will 
    either be in tumbling state, or have satellites.
    Automatic modeling procedures should be able to detect such cases,
    triggering detailed analysis of these peculiar targets.

The rising interest of private companies for near-Earth objects as mining resources will inevitably lead to further development of analysis of remote-sensing data, building dedicated ground-based or space telescopes, and eventually moving from asteroids surface into their interiors.

\bigskip

\textbf{Acknowledgments.} The work of J\v{D} was supported by the grant GACR P209/10/0537 of the Czech Science Foundation. MD acknowledges support from the grant ANR-11-BS56-008 SHOCKS of the French National Research Agency. 

\newcommand{\SortNoop}[1]{}

 
\end{document}